\documentclass[preprint,prb,aps,eqsecnum]{revtex4}
\usepackage{graphicx}
\usepackage{bm}
\usepackage{epsf}

\begin{document}

\def\bb{\begin{eqnarray}}
\def\ee{\end{eqnarray}}
\title
{X-ray resonant magnetic scattering from structurally and
magnetically rough interfaces in multilayered systems\\
I. Specular reflectivity}

\author{D. R. Lee}
        \email{drlee@aps.anl.gov}
\author{D. Haskel}
\author{Y. Choi}
    \altaffiliation[Also at ]{Department of Materials Science and Engineering,
                              Northwestern University, Evanston, IL 60208}
\author{J. C. Lang, S. A. Stepanov}
\affiliation{Advanced Photon Source, Argonne National Laboratory, Argonne,
Illinois 60439}

\author{G. Srajer}
\affiliation{Advanced Photon Source, Argonne National Laboratory, Argonne,
Illinois 60439}

\author{S. K. Sinha}
\affiliation{Department of Physics, University of California, San Diego, La
Jolla, CA 92093, and\\ Los Alamos National Laboratory, Los Alamos, NM 87545}

\date{\today}

\begin{abstract}
The theoretical formulation of x-ray resonant magnetic scattering from
rough surfaces and interfaces is given for specular reflectivity. 
A general expression is derived for both structurally and magnetically rough
interfaces in the distorted-wave Born approximation (DWBA) as the framework
of the theory. 
For this purpose, we have defined a ``structural'' and a ``magnetic'' interface
to represent the actual interfaces.
A generalization of the well-known Nevot-Croce formula for 
specular reflectivity is obtained for the case of a single rough magnetic interface
using the self-consistent method. Finally, the results are generalized to the
case of multiple interfaces, as in the case of thin films or multilayers.
Theoretical calculations for each of the cases are illustrated with numerical
examples and compared with experimental results of magnetic reflectivity 
from a Gd/Fe multilayer.
\end{abstract}
\maketitle


\section{INTRODUCTION}
X-ray reflectivity and offspecular diffuse scattering methods have been widely
applied over the last decade to characterize the morphology of rough surfaces
and interfaces, particularly with the availability of sources of
ever-increasing brilliance for x-ray radiation. 
Similar techniques using neutron beams have also become widespread, 
particularly for the study of magnetic multilayers. 
In the case of x-rays, however, element-specific information
regarding the magnetic structure can be readily obtained by tuning the photon
energy to that of an L-edge (in the case of transition or rare-earth
metals)\cite{gibbs88,kao94}  or of an M-edge (in the case of
actinides).\cite{isaacs89,mcwhan90} 
The resonant enhancement of the scattering
by magnetic atoms at such energies can result in a large enough signal to be
comparable to the dominant charge scattering. 
Resonant x-ray scattering at the K-edges of transition metals\cite{namikawa85} 
has also been used to obtain information about the magnetic structure, 
although the enhancement is not as large. 
Resonant magnetic scattering corresponds to the
real part of the scattering amplitude, while the (absorptive) imaginary part
gives rise to x-ray magnetic circular dichroism (XMCD), which has been used to
obtain the values of spin and orbital moments in ferromagnetic materials.
Detailed descriptions of the formalism for the interaction of x-rays with
magnetically polarized atoms have been given in the
literature,\cite{hannon88,blume88,hamrick,hill96,lovesey} from which a
complete description of magneto-optic phenomena in the x-ray region can be
obtained and applied.

Several resonant x-ray specular reflectivity experiments have been performed to 
obtain the magnetization within the layers of magnetic
multilayers.\cite{kao94,tonnerre95,sacchi98,ishimatsu99,haskel01} 
The analysis of these results has generally used recursive matrix techniques
developed for magneto-optics in the case of resonant x-ray
reflectivity.\cite{stepanov} 
In general, roughness at the interfaces has been
ignored or taken into account in an ad-hoc manner. 
In principle, representing
roughness in terms of a graded magnetization at the interface and using
slicing methods could enable one to calculate the effect of magnetic
roughness on specular reflectivity at the expense of considerable
computational effort. 
R\"{o}hlsberger has developed a matrix formalism
(originally developed for nuclear resonant x-ray reflectivity) from which
specular reflectivity incorporating roughness can be
calculated.\cite{rohlsbeger99}
It was not considered in his paper, however, that the magnetic interfaces
can have different roughnesses from the structural (chemical) ones.
In this paper, we define separately a structural and a magnetic interface
to represent the actual interfaces and present analytical formulae 
taking into account both interface roughnesses, 
which provide much faster computational method than the slicing methods
and show good agreement with established formulae for chemical interface
roughness.

Methods were developed earlier to calculate analytically the specular
component of the charge scattering of x-rays by rough surfaces and interfaces
using the Born approximation (BA) and the distorted-wave Born approximation
(DWBA).\cite{sinha88,holy94} 
The BA results were extended to magnetic interfaces 
in an earlier publication\cite{osgood} and have
already been applied to interpreting x-ray resonant magnetic specular reflectivity
measurements from magnetic multilayers.\cite{haskel01}
However, the BA or the kinematical approximation breaks down in the vicinity
of the critical angle and below, since it neglects the x-ray refraction.
On the other hand, the DWBA takes account of dynamical effects, such as
multiple scattering and the x-ray refraction, which become significant
for smaller angles close to the critical angle and even for greater angles
at the resonant energies or with soft x-rays. 
We present here the generalization of the DWBA 
to the case of resonant magnetic x-ray reflectivity from rough magnetic
surfaces or interfaces. 
The principal complication is, however, that we now have to
deal with a tensor (rather than scalar) scattering length, or equivalently an
anisotropic refractive index for x-rays.\cite{stepanov} 
This leads in general to two transmitted and two reflected waves 
at each interface for arbitrary polarization, 
which complicates the DWBA formalism.

The plan of this paper is as follows. 
In Sec. II, we discuss a  simple conceptual model for a magnetic interface 
and its relationship to the chemical (i.e., structural) interface 
and define the appropriate magnetic roughness parameters.
In Sec. III, we discuss the (known)
scattering amplitudes for resonant x-ray scattering and their relationship to
the dielectric susceptibility to be used in the DWBA. 
In Sec. IV, we present the derivation of the scattering in the DWBA 
for a single interface with both structural and magnetic roughnesses. 
In Secs. V and VI, we derive the formulae for specular reflectivity 
from a magnetic interface using the self-consistent method in the framework 
of the DWBA and discuss numerical results. 
Finally, in Secs. VII-IX, we discuss the extension of the formalism
to the case of the specular reflectivity from magnetic multilayers and present
some numerical results with experimental data from a Gd/Fe multilayer. 
In the following paper,\cite{paperII} we derive the formulae for the diffuse
(off-specular) scattering from magnetic interfaces in both the BA and the DWBA.


\section{MODEL FOR MAGNETIC INTERFACE} 
Consider an interface between a ferromagnetic medium and a nonmagnetic medium 
(which could also be free space). 
Due to the roughness of this interface, the magnetic moments near the interface 
will find themselves in anisotropy and exhange fields, which fluctuate spatially 
(see Fig. 1).

This will produce disorder relative to the preferred ferromagnetic alignment 
within the magnetic medium.
A similar situation can arise at an interface between a ferromagnetic medium (FM) 
and an antiferromagnetic medium (AFM), where there is a strong antiferromagnetic 
coupling between spins in the FM and the AFM.
Random steps will then produce frustration in the vicinity of the interface, 
resulting in random disordering of the magnetic moments near the interface.
Clearly in general correlation will exist between the height 
fluctuations of the chemical interface and the fluctuations of the spins, 
but a quantitative formalism to account for this in detail 
has not yet been developed.
We make here the simplifying assumption that the ferromagnetic moments 
near the interface (or at least their components in the direction of 
the ferromagnetic moments deep within the FM layer, i.e., the direction of 
average magnetization ${\bf \hat{M}}$) are cut off at a mathematical interface,
which we call the magnetic interface 
and which may not coincide with the chemical interface, 
either in its height fluctuations or over its average position, 
e.g., if a magnetic ``dead layer'' exists between the two interfaces (see Fig. 1).  
The disorder near the interface is thus represented by height fluctuations 
of this magnetic interface.
The basis for this assumption, which is admittedly crude, is that the short (i.e., 
atomic) length-scale fluctuations of the moments away from the direction of 
the average magnetization give rise to diffuse scattering at fairly large scattering 
wave vectors, whereas we are dealing here with scattering
at a small wave vector ${\bf q}$, which represent the relatively slow variations 
of the average magnetization density.
The actual interface can be then considered as really composed of two interfaces,
a chemical interface and a magnetic interface, each with their own average height, 
roughness, and correlation length, and, importantly, in general possessing
correlated height fluctuations.


\section{RESONANT MAGNETIC X-RAY SCATTERING Amplitude}

The amplitude for resonant magnetic scattering of x-rays has been derived by
Hannon et al.,\cite{hannon88} and a discussion of the general formalism may
be found in the review by Hill and McMorrow.\cite{hill96} 
There are two cases of practical importance, namely dipole and quadrupole resonances. 
We shall restrict ourselves here to the most commonly used dipole resonance, 
which is related to the L-edges of transition metals and rare-earth atoms. 
The tensor amplitude for scattering $f_{\alpha\beta}$ from a magnetic atom is given by
\bb\label{eq:f_tensor}
\sum_{\alpha\beta} e^{\ast}_{f\alpha}f_{\alpha\beta} e_{i\beta} &=&
\left[ f_0 +
\frac{3\lambda}{8\pi}(F_{11}+F_{1-1})\right]
\Bigl({\bf\hat{e}}^{\ast}_f\cdot{\bf\hat{e}}_i\Bigr) \nonumber \\
&-& i \frac{3\lambda}{8\pi}(F_{11}-F_{1-1})
           \Bigl({\bf\hat{e}}^{\ast}_f\times{\bf\hat{e}}_i\Bigr)\cdot{\bf\hat{M}}
        \nonumber \\
&+& \frac{3\lambda}{8\pi}(2F_{10}-F_{11}-F_{1-1})
         \Bigl({\bf\hat{e}}^{\ast}_f\cdot{\bf\hat{M}}\Bigr)
	\Bigl({\bf\hat{e}}_i\cdot{\bf\hat{M}}\Bigr),
\ee
where ${\bf\hat{e}}_i$, ${\bf\hat{e}}_f$ are, respectively,
the unit photon polarization
vectors for the incident and scattered waves, ${\bf\hat{M}}$ is a unit vector in
the direction of the magnetic moment of the atom, $\lambda$ is the x-ray
photon wavelength, $f_0$ is the usual Thomson (charge) scattering amplitude
[$f_0 = -r_0(Z+f^{\prime}-i f^{\prime\prime}$)], 
where $r_0$ is the Thomson scattering length ($e^2/mc^2$),  
$Z$ is the atomic number, 
$f^{\prime}(<0)$ and $f^{\prime\prime}(>0)$ are the real and imaginary
non-resonant dispersion corrections.
$F_{LM}$ is the resonant scattering
amplitude, as defined in Ref. \onlinecite{hannon88},
and has the resonant denominator $E_{\rm res}-E-i\Gamma/2$, which
provides the resonance when the photon energy $E$ is tuned to the resonant
energy $E_{\rm res}$ close to the absorption edges.
The lifetime of the resonance $\Gamma$ is typically $1-10$ eV, 
so that the necessary energy resolution is easily achivable at
synchrotron radiation beamlines. 
(We assumed that ${\bf q}$, the wave-vector transfer, 
is small enough here that the atomic form factor can be taken as unity.) 
Equation (\ref{eq:f_tensor}) has both real and imaginary
(i.e., absorptive) components. 
The latter gives rise to the well-known phenomenon of x-ray magnetic circular 
or linear dichroism, whereas the real part gives rise to the scattering. 
Equation (\ref{eq:f_tensor}) yields
\bb \label{eq:f_ABC}
f_{\alpha\beta} = A \delta_{\alpha\beta} - i B \sum_{\gamma}
\epsilon_{\alpha\beta\gamma} M_{\gamma} + C M_{\alpha} M_{\beta},
\ee
where
\bb \label{eq:ABC}
A &=& f_0 + \frac{3\lambda}{8\pi}(F_{11}+F_{1-1}), \nonumber \\
B &=& \frac{3\lambda}{8\pi}(F_{11}-F_{1-1}),\nonumber \\
C &=& \frac{3\lambda}{8\pi}(2F_{10}-F_{11}-F_{1-1}),
\ee
and $\alpha$, $\beta$ denote Cartesian components, and
$\epsilon_{\alpha\beta\gamma}$ is the antisymmetric Levi-Civita symbol
($\epsilon_{xyz} =\epsilon_{yzx}= \epsilon_{zxy}=1$,
$\epsilon_{xzy}=\epsilon_{yxz}= \epsilon_{zyx}=-1$, all other
$\epsilon_{\alpha\beta\gamma}=0$). 
The dielectric susceptibility of a resonant magnetic medium is given by
\bb \label{eq:chi_f}
\chi^{\rm resonant}_{\alpha\beta}({\bf r}) =
         \frac{4\pi}{k_0^2}n_m({\bf r})f_{\alpha\beta}({\bf r}),
\ee
where $k_0 = 2\pi/\lambda$, $n_m({\bf r})$ is the local number density of
resonant magnetic atoms, and the variation of $f_{\alpha\beta}({\bf r})$ with
${\bf r}$ reflects the possible positional dependence of the direction of
magnetization ${\bf M}$. 
The total dielectric susceptibility is given by
\bb \label{eq:chi_tot}
\chi_{\alpha\beta}({\bf r}) &=& \frac{4\pi}{k_0^2}\biggl[
     \Bigl\{-\rho_0 ({\bf r}) r_0 + A n_m({\bf r})\Bigr\} \delta_{\alpha\beta}
                                                                  \nonumber \\
                           &-& i B n_m({\bf r}) \sum_{\gamma}
                               \epsilon_{\alpha\beta\gamma}M_{\gamma}({\bf r})
                + C n_m({\bf r})M_{\alpha}({\bf r})M_{\beta}({\bf r}) \biggr],
\ee
where $\rho_0({\bf r})$ represents the electron number density arising from
all the other nonresonant atoms in the medium modified by their anomalous
dispersion corrections when necessary. 
Using the constitutive relationship between the local dielectric constant tensor
$\epsilon_{\alpha\beta}({\bf r})$ and $\chi_{\alpha\beta}({\bf r})$,
\bb \label{eq:dielectric}
\epsilon_{\alpha\beta}({\bf r})
                         = \delta_{\alpha\beta} + \chi_{\alpha\beta}({\bf r}).
\ee
We note that the magnetization gives the dielectric tensor the same symmetry
as in conventional magneto-optic theory, namely an antisymmetric component
linear in the magnetization.


\section{The Distorted-Wave Born Approximation for a Single Magnetic Interface}

The results for specular reflectivity in the Born approximation (BA) have been
derived in Ref. \onlinecite{osgood} and will be also summarized briefly 
in connection with the cross section in the following paper.\cite{paperII}  
Here we discuss the scattering in terms of the distorted-wave Born approximation (DWBA).
While this is more complicated algebraically, it provides a better description 
than the simple kinematical approximation or BA
in the vicinity of regions where total reflection or Bragg scattering occurs.
This treatment is a generalization of that used in Ref. \onlinecite{sinha88}
for charge scattering. 
The wave equation for electromagnetic waves propagating in an anisotropic medium 
with a dielectric susceptibility tensor given by Eq. (\ref{eq:chi_tot}) 
may be written as
\bb \label{eq:wave_eq}
\sum_{\beta} \left[ (\nabla^2 + k^2_0 ) \delta_{\alpha\beta} -
\nabla_{\alpha}\nabla_{\beta} + k_0^2 \chi_{\alpha\beta}\right]
E_{\beta}({\bf r}) = 0,~~ (\alpha, \beta = x, y, z),
\ee
where ${\bf E}({\bf r})$ is the electric field vector.

\begin{figure}
\epsfxsize=12cm
\centerline{\epsffile{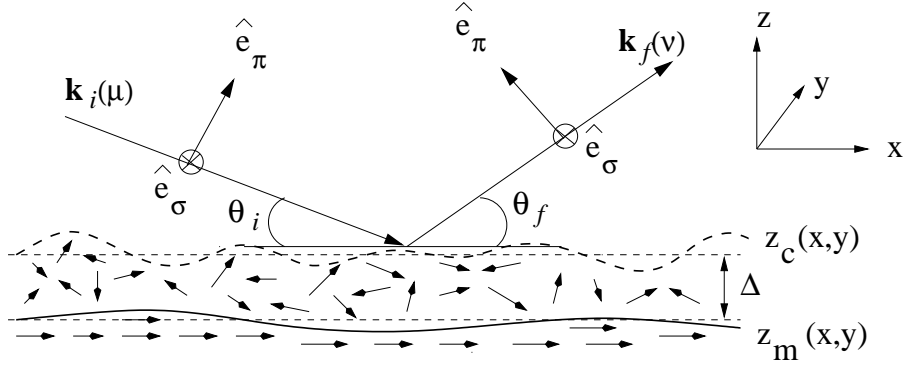}}
\caption{\label{fig-scatt-geo}
Schematic of scattering geometry and sketch of the chemical (or structural) 
($z_c (x,y)$) and magnetic ($z_m (x,y)$) interfaces, 
which can be separated from one another by an average amount $\Delta$.
Grazing angles of incidence ($\theta_i$) and scattering ($\theta_f$),
the wave vectors ${\bf k}_i$ and ${\bf k}_f$, and the photon polarization vectors 
of incidence (${\bf\hat{e}}_{\mu=\sigma,\pi}$) and 
scattering (${\bf\hat{e}}_{\nu=\sigma,\pi}$) are illustrated.
Small arrows represent the possible orientations of the magnetic moments
around magnetic interfaces.  
}
\end{figure}

Consider a wave incident, as in Fig. \ref{fig-scatt-geo} 
with wave vector ${\bf k}_i$ in
the $(x,z)$ plane ($k_{i,y} = 0$) and polarization $\mu~(\mu=\sigma$ or $\pi$),
from a nonmagnetic (isotropic) medium for which $\chi_{\alpha\beta} =\chi_0
\delta_{\alpha\beta}$ onto a smooth interface at $z=0$ with a magnetic medium,
for which $\chi_{\alpha\beta}$ is constant for $z<0$.

Let us write for $z<0$
\bb \label{eq:chi_12}
\chi_{\alpha\beta} = \chi_1 \delta_{\alpha\beta} + \chi^{(2)}_{\alpha\beta},
\ee
where the term $\chi^{(2)}_{\alpha\beta}$ is the part that specifically
depends on the magnetization ${\bf M}$, as defined in Eq. (\ref{eq:chi_tot}). 
The incident wave (chosen for convenience with unit amplitude) may be written as
\bb\label{eq:Ei}
{\bf E}^i_{\mu}({\bf r}) = {\bf\hat{e}}_{\mu} e^{i {\bf k}_i\cdot{\bf r}}.
\ee
This incident wave will in general give rise to two specularly reflected
waves (where the index $\mu$ refers to $\sigma$ or $\pi$ polarization) 
and two transmitted (refracted) waves in the magnetic medium. 
The complete solution for the electric field in
the case of the smooth magnetic interface is then given by
\bb\label{eq:E_ki}
{\bf E}_{({\bf k}_i,\mu)}({\bf r}) &=&
{\bf\hat{e}}_{\mu} e^{i{\bf k}_i\cdot{\bf r}} + \sum_{\nu=\sigma,\pi}
R^{(0)}_{\nu\mu}({\bf k}_i) {\bf\hat{e}}_{\nu} e^{i{\bf k}_i^{r}\cdot{\bf r}}, ~~z>0,
\nonumber \\
&=& \sum_{j=1,2} T^{(0)}_{j\mu}({\bf k}_i){\bf\hat{e}}_j
e^{i{\bf k}_i^{t}(j)\cdot{\bf r}}, ~~z<0,
\ee
where ${\bf k}_i^{r}$ is the specularly reflected wave vector in the
nonmagnetic medium, $\nu$ denotes the polarization of the appropriate
reflected component, the index $j(=1,2)$ defines the component of the
transmitted wave in the magnetic resonant medium with polarization ${\bf\hat{e}}_j$
(${\bf\hat{e}}_{j=1,2} = {\bf\hat{e}}^{(1)}$ and  ${\bf\hat{e}}^{(2)}$, respectively,
as defined in Appendix A), and
${\bf k}_i^{t}(j)$ the appropriate wave vector for that transmitted wave.
The polarization vectors ${\bf\hat{e}}$ may be real or complex 
allowing for linear or elliptically polarized waves. 
We denote such states in Eq. (\ref{eq:E_ki}) 
quantum-mechanically by $|{\bf k}_i, \mu >$.

$R^{(0)}_{\nu\mu}$ and $T^{(0)}_{j\mu}$ denote 
the appropriate reflection and transmission coefficients for the smooth surface
and are expressed in terms of $2\times 2$ matrices using the polarization bases
for the incident and reflected (or transmitted) waves. 
The polarization basis is given by (${\bf\hat{e}}_{\sigma}$, ${\bf\hat{e}}_{\pi}$),
as shown in Fig. 1, for the waves in the nonmagnetic medium and 
(${\bf\hat{e}}^{(1)}$, ${\bf\hat{e}}^{(2)}$), as defined in Appendix A, 
for those in the magnetic resonant medium, respectively. 
The convention in which the polarization state of the reflected (or transmitted) wave 
precedes that of the incident wave is used 
for the subscripts in $R^{(0)}_{\nu\mu}$ and $T^{(0)}_{j\mu}$, 
and the Greek and Roman letters are used for the polarization states in the
nonmagnetic and magnetic medium, respectively. 
The explicit expressions of $R^{(0)}_{\nu\mu}$ and $T^{(0)}_{j\mu}$
for small angles of incidence and 
small amplitudes of the dielectric susceptibility and
for special directions of the polarization and magnetization (i.e., 
${\bf M}\parallel{\bf\hat{x}}$ as shown in Fig. 1) are
given in Appendix A.

We should mention, however, that these specific conditions considered
in Appendix A (and also in all other appendices) are reasonably satisfied
for hard- and medium-energy x-rays and also for soft x-rays around
transition-metal L-edges with small angles (i.e., when 
$\theta_i^2 \ll 1$ for the incidence angle $\theta_i$).
We should also mention that, even when ${\bf M}$ is not parallel to 
the ${\bf\hat{x}}$-axis in Fig. 1, the expressions derived in the appendices 
can be still applied by considering only the $x$-component of the magnetization
vector ${\bf M}$.
This is because the $y$- and $z$-components of ${\bf M}$ contribute negligibly
to the scattering in comparison with
with the dominant factor $B=(3\lambda/8\pi)(F_{11}-F_{1-1})$ in Eq. (\ref{eq:f_ABC})
at small angles\cite{stepanov} when $|F_{11}-F_{1-1}|\gg |2F_{10}-F_{11}-F_{1-1}|$,
which is generally satisfied for transition-metal and rare-earth 
L-edges.\cite{hamrick}

We note that the continuity of the fields parallel to the interface requires that
\bb\label{eq:contin_ki}
({\bf k}_i)_{\parallel} = ({\bf k}_i^{r})_{\parallel}
                        = \Bigl({\bf k}_i^{t}(j)\Bigr)_{\parallel} ,
\ee
where $( )_{\parallel}$ denotes the vector component parallel to the interface.

We now discuss the structurally and magnetically rough interface. 
For this purpose we shall assume that the average height (along $z$) of 
the structural and magnetic interfaces is the same, i.e., 
we ignore the presence of a magnetic dead layer. 
This may be treated within the DWBA as simply another
nonmagnetic layer and thus discussed within the formalism for treating
multilayers as discussed in Section VII. 
We can write
\bb\label{eq:chi_perturb}
\chi_{\alpha\beta}({\bf r}) = \chi_{\alpha\beta}^{(0)}({\bf r})+
\Delta_{\alpha\beta}^c({\bf r}) + \Delta_{\alpha\beta}^m({\bf r}),
\ee
where
\bb\label{eq:chi0}
\chi_{\alpha\beta}^{(0)}({\bf r}) &=& \chi_0 \delta_{\alpha\beta}, ~~z>0
\nonumber \\
&=& \chi_1 \delta_{\alpha\beta} + \chi^{(2)}_{\alpha\beta}, ~~z<0,
\ee
\bb\label{eq:perturb_c}
\Delta_{\alpha\beta}^c({\bf r})&=& (\chi_1 - \chi_0) \delta_{\alpha\beta},
~~{\rm for}~~ 0<z<\delta z_c(x,y)~~ {\rm if}~~\delta z_c(x,y)>0  \nonumber \\
&=& -(\chi_1 - \chi_0)\delta_{\alpha\beta},
     ~~{\rm for}~~ \delta z_c(x,y)<z<0~~{\rm if}~~\delta z_c(x,y)<0
                                                                 \nonumber \\
&=& 0 ~~~~{\rm elsewhere},
\ee
and
\bb\label{eq:perturb_m}
\Delta_{\alpha\beta}^m({\bf r})&=& \chi_{\alpha\beta}^{(2)}, ~~{\rm for}~~
0<z<\delta z_m(x,y)~~ {\rm if}~~\delta z_m(x,y)>0  \nonumber \\
&=& -\chi_{\alpha\beta}^{(2)}, ~~{\rm for}~~ \delta z_m(x,y)<z<0~~
{\rm if}~~\delta z_m(x,y)<0  \nonumber   \\
&=& 0 ~~~~{\rm elsewhere},
\ee
$\delta z_c(x,y)$ and $\delta z_m(x,y)$ define the structural (chemical) and magnetic
interfaces, respectively.

\begin{figure}
\epsfxsize=10cm
\centerline{\epsffile{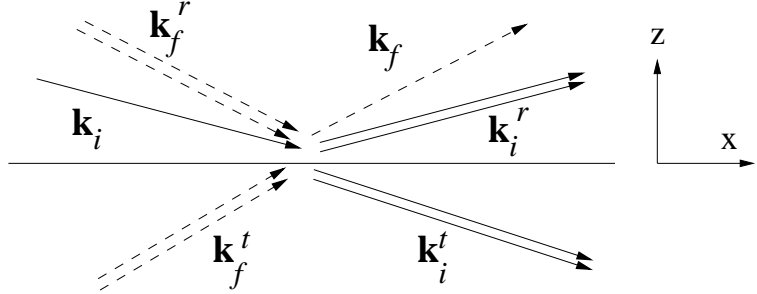}}
\caption{\label{fig-dwba-geo}
Schematic of an ideal interface with undisturbed states ${\bf E}({\bf k}_i)$
and ${\bf E}^T(-{\bf k}_f)$.
Note two possible waves for each of the reflected and transmitted wave vectors.}
\end{figure}

We may also define the time-reversed function corresponding to a wave
incident on the interface with vector $(-{\bf k}_f)$ and polarization $\nu$ as
\bb\label{eq:E_kf}
{\bf E}^T_{(-{\bf k}_f, \nu)}({\bf r}) &=&
{\bf\hat{e}}_{\nu}e^{i {\bf k}_f^{\ast}\cdot{\bf r}}
+ \sum_{\lambda=\sigma,\pi} R^{(0)\ast}_{\lambda\nu}(-{\bf k}_f)
 {\bf\hat{e}}_{\lambda} e^{i {\bf k}_f^{r\ast}\cdot{\bf r}} ,~~z>0      \nonumber \\
&=& \sum_{j=1,2} T^{(0)\ast}_{j\nu} (-{\bf k}_f) {\bf\hat{e}}_j
 e^{i {\bf k}_f^{t\ast}(j)\cdot{\bf r}}, ~~z<0,
\ee
where $(-{\bf k}^r_f)$ is the wave vector of the wave specularly reflected
from $(-{\bf k}_f)$, and $\left(-{\bf k}^t_f(j)\right)$ is the wave vector 
of one of the
two transmitted waves in the medium emanating from $(-{\bf k}_f)$ incident on
the surface, as shown in Fig. \ref{fig-dwba-geo}.
Note that, for consistency with the conventions used in Eq. (\ref{eq:E_ki}),
the polarization vectors in Eq. (\ref{eq:E_kf}) are defined 
in the ordinary coordinate system where their phases are considered along the left-to-right
direction in Fig. 1.
Otherwise, the polarization vectors in Eq. (\ref{eq:E_kf}) should be replaced by
their complex conjugates. 

We have also the conditions
\bb\label{eq:contin_kf}
({\bf k}_f)_{\parallel} = ({\bf k}_f^r)_{\parallel}
                = \left({\bf k}_f^t(j)\right)_{\parallel}.
\ee
The DWBA then yields the differential cross section for scattering by the
rough interface from $({\bf k}_i, \mu)$ to $({\bf k}_f, \nu)$ as
\bb\label{eq:cross_section}
\frac{d\sigma}{d\Omega} = \frac{1}{16\pi^2} \Bigl< |{\cal T}^{fi}|^2 \Bigr>,
\ee
where  ${\cal T}^{fi}=<{\bf k}_f, \nu|{\cal T}| {\bf k}_i, \mu >$ is 
the scattering matrix element, and $\Bigl<...\Bigr>$ in
Eq. (\ref{eq:cross_section}) denotes a statistical averaging over random 
fluctuations at the interface. 
Following Ref. \onlinecite{sinha88}, we split the cross section into two parts:
\bb\label{eq:cross_section_dwba}
\frac{d\sigma}{d\Omega} = \frac{1}{16\pi^2}
         \Bigl | \Bigl<{\cal T}^{fi}\Bigr> \Bigr|^2
	+ \frac{1}{16\pi^2}
         \biggl [ \Bigl<|{\cal T}^{fi}|^2 \Bigr> - 
	   \left|\Bigl<{\cal T}^{fi}\Bigr>\right|^2 \biggr].
\ee
The first term in Eq. (\ref{eq:cross_section_dwba}) represents the
coherent (specular) part of the scattering, which corresponds to a statistical
averaging of the scattering amplitude, and the second term corresponds to
the incoherent (diffuse) scattering. 
In this paper, we shall deal with the first term
only, while the diffuse scattering will be addressed 
in the following paper.\cite{paperII}

The DWBA consists of approximating the scattering matrix element by the expression
\bb\label{eq:scatt_mat_element}
<{\bf k}_f, \nu|{\cal T}|{\bf k}_i,\mu>  &=& k_0^2<-{\bf k}^T_f, \nu | {\bm\chi}^{(0)}
| {\bf E}^i_{\mu}({\bf r})> \nonumber \\
&+& k_0^2<-{\bf k}^T_f, \nu | {\bm \Delta}^{c} | {\bf k}_i, \mu>
+ k_0^2<-{\bf k}^T_f, \nu | {\bm \Delta}^{m} | {\bf k}_i, \mu>.
\ee
Here $|{\bf E}^i_{\mu}({\bf r})>$ denotes the ``pure'' incoming wave in Eq. (\ref{eq:Ei}),
$|-{\bf k}^T_f, \nu>$ denotes the state in Eq. (\ref{eq:E_kf}), 
and the matrix element involves dot products of the tensor operators 
${\bm\chi}^{(0)}$, ${\bm\Delta}^c$, and ${\bm \Delta}^m$ 
with the vector fields $<-{\bf k}^T_f, \nu|$ and $<{\bf k}_i, \mu|$. 
While ${\bm\chi}^{(0)}$ represents an ideal system
with a smooth interface, ${\bm \Delta}^c$ and ${\bm \Delta}^m$ are
perturbations on ${\bm\chi}^{(0)}$ due to interface roughnesses.

For the smooth surface, only the first tensor is nonvanishing, and, following
Ref.~\onlinecite{sinha88}, we can show from Eqs. (\ref{eq:Ei}) and 
(\ref{eq:E_kf}) that
\bb\label{eq:chi0_mat}
k_0^2<-{\bf k}^T_f, \nu | \bm{\chi}^{(0)} | {\bf E}^i_{\mu}({\bf r})>
&=& i{\cal A} k_0^2\delta_{k_{ix}k_{fx}} \delta_{k_{iy}k_{fy}}           \nonumber \\
&\times& \sum_j T^{(0)}_{j\nu}(-{\bf k}_f)
	\sum_{\alpha\beta} e^{\ast}_{j\alpha}
		(\chi_1 \delta_{\alpha\beta} + \chi_{\alpha\beta}^{(2)})
                 e_{\mu\beta} \nonumber \\
&\times& \int^{0}_{-\infty} dz e^{-i ( k^t_{fz}(j)-k_{iz})z },    \nonumber \\
&=& 2i {\cal A} k_{iz} R^{(0)}_{\nu\mu}({\bf k}_i) \delta_{k_{ix}k_{fx}}
                                                \delta_{k_{iy}k_{fy}},
\ee
where ${\cal A}$ is the illuminated surface area, and 
$R^{(0)}_{\nu\mu}({\bf k}_i)$ is the reflection coefficient for the smooth
surface, as defined in Eq. (\ref{eq:E_ki}).
The details of Eq. (\ref{eq:chi0_mat}) are presented in Appendix B.  
By comparison with Eq. (\ref{eq:chi0_mat}) for the smooth surface, 
the scattering matrix element for the rough surface
in Eq. (\ref{eq:scatt_mat_element}) can be analogously defined by
\bb\label{eq:T_R}
<{\bf k}_f, \nu|{\cal T}|{\bf k}_i,\mu>  = 
	2i {\cal A} k_{iz} R_{\nu\mu}({\bf k}_i) \delta_{k_{ix}k_{fx}} \delta_{k_{iy}k_{fy}}, 
\ee
where $R_{\nu\mu}({\bf k}_i)$ denotes the reflection coefficient for the rough surface.

On the other hand, for the reverse case where a wave is incident 
from a resonant magnetic medium to a
nonmagnetic (isotropic) medium, similarly to Eq. (\ref{eq:chi0_mat}),
the scattering matrix element for the smooth surface can be shown to be 
\bb \label{eq:chi0_mat_rn}
k_0^2<-{\bf k}^T_f, j^{\prime} |\bm{\chi}^{(0)} | {\bf k}_i, j > 
= 4i{\cal A} k_{iz}(j) R^{(0)}_{j^{\prime} j}( {\bf k}_i )
	\delta_{k_{ix}k_{fx}} \delta_{k_{iy}k_{fy}}, 
\ee
where the incoming wave from the resonant magnetic medium $| {\bf k}_i, j > $
is used instead of the ``pure'' incoming wave from the vacuum ${\bf E}^i_{\mu}({\bf r})$ 
in Eq. (\ref{eq:Ei}). 
The use of Eqs. (\ref{eq:chi0_mat}) and (\ref{eq:chi0_mat_rn}) in Eqs.
(\ref{eq:scatt_mat_element}) and (\ref{eq:cross_section}) in the case of
the smooth surface and the derivation of the corresponding reflectivity
in the usual manner, as discussed in Ref. \onlinecite{sinha88},
shows that Eqs. (\ref{eq:chi0_mat}) and (\ref{eq:chi0_mat_rn}) must be identically true.
Similarly to Eqs. (\ref{eq:chi0_mat}) and (\ref{eq:T_R}), 
the scattering matrix element for the rough surface
between reversed layers can be also defined by analogy 
from Eq. (\ref{eq:chi0_mat_rn}) as
\bb\label{eq:T_R_rn}
<{\bf k}_f, j^{\prime}|{\cal T}|{\bf k}_i, j>  = 
	4i {\cal A} k_{iz}(j) R_{j^{\prime} j}({\bf k}_i) 
	\delta_{k_{ix}k_{fx}} \delta_{k_{iy}k_{fy}}, 
\ee
where $R_{j^{\prime} j}({\bf k}_i)$ denotes 
the reflection coefficient for the rough surface between reversed layers.


\section{Reflection and transmission coefficients using the Self-Consistent Method}

To calculate specular reflectivity, we make an approximation in the
spirit of Nevot and Croce.\cite{nevot} 
To evaluate the matrix elements in Eq. (\ref{eq:scatt_mat_element}) involving 
$\Delta_{\alpha\beta}^c$ and $\Delta_{\alpha\beta}^m$, 
we assume for ${\bf E}({\bf k}_i, \mu)$ in Eq. (\ref{eq:E_ki}) 
the functional form for $z>0$ analytically continued for $z<0$, while
for the time-reversed state ${\bf E}^{T}(-{\bf k}_f, \nu)$ in Eq. (\ref{eq:E_kf}) 
the functional form for $z<0$ analytically continued to $z>0$. 
Then, bearing in mind that for specular reflectivity ${\bf k}_f = {\bf k}_i^r$ 
and using Eq. (\ref{eq:contin_ki}), we obtain
for the statistically averaged amplitude $\Bigl<{\cal T}^{fi}\Bigr>$:
\bb\label{eq:chi_cm_mat}
&~&
\Bigl< k_0^2<-{\bf k}^T_f, \nu|{\bf\Delta}^{c,m}|{\bf k}_i, \mu > \Bigr>
= i {\cal A} k_0^2 \sum_{j=1,2} T^{(0)}_{j\nu}(-{\bf k}_f) ~~~~~~~~~~~~~  \nonumber \\
&~&~~~~~~~~~~~~~\times \Biggl[\sum_{\alpha\beta}
\frac{e_{j\alpha}^{\ast} \Delta^{c,m}_{\alpha\beta}e_{\mu\beta}}{q_{1z}(j)}
                \biggl[\Bigl<e^{-i q_{1z}(j)\delta z_{c,m}(x,y)}\Bigr> -1 \biggr]  
		     \nonumber \\
&~&~~~~~~~~~~~~~
 + \sum_{\lambda=\sigma,\pi} R^{(0)}_{\lambda\mu}({\bf k}_i) \sum_{\alpha\beta}
\frac{e_{j\alpha}^{\ast} \Delta^{c,m}_{\alpha\beta}e_{\lambda\beta}}{q_{2z}(j)}
                \biggl[\Bigl<e^{-i q_{2z}(j)\delta z_{c,m}(x,y)}\Bigr> -1 \biggr]
\Biggr],
\ee
where
\bb\label{eq:qz_12}
q_{1z}(j) = k_{fz}^t(j)-k_{iz},~~~
q_{2z}(j) = k_{fz}^t(j)-k_{iz}^r,
\ee
and $\Delta^{c,m}_{\alpha\beta}$ is the value defined for $0<z<\delta z_{c,m}$
in Eqs. (\ref{eq:perturb_c}) and (\ref{eq:perturb_m}). 
From Eqs. (\ref{eq:chi0_mat})-(\ref{eq:T_R}) and (\ref{eq:chi_cm_mat}), we see
that, at the specular condition, we can write Eq. (\ref{eq:scatt_mat_element}) as
\bb\label{eq:RUV_mat}
R_{\nu\mu} = R^{(0)}_{\nu\mu} + U_{\nu\mu} + \sum_{\lambda}
V_{\nu\lambda}R^{(0)}_{\lambda\mu},
\ee
where
\bb\label{eq:U_mat}
U_{\nu\mu} &=& \sum_{j=1,2} \frac{ T^{(0)}_{j\nu}(-{\bf k}_f)}{2k_{iz}}
\frac{k_0^2}{q_{1z}(j)}
\biggl[ (\chi_1 - \chi_0) \sum_{\alpha} e_{j\alpha}^{\ast} e_{\mu\alpha}
\bigl[ e^{-\frac{1}{2}q_{1z}^2(j)\sigma_c^2} - 1 \bigr]  \nonumber \\
&~&~~~~~~~~~~+ \sum_{\alpha\beta}e_{j\alpha}^{\ast} \chi^{(2)}_{\alpha\beta}
e_{\mu\beta} \bigl[ e^{-\frac{1}{2}q_{1z}^2(j)\sigma_m^2} - 1 \bigr] \biggr],
\ee
and replacing $q_{1z}$, $e_{\mu}$ in $U_{\nu\mu}$ by $q_{2z}$,
$e_{\lambda}$ produces $V_{\mu\lambda}$. 
Here we made the
customary Gaussian approximation for the height fluctuations $\delta z_{c,m}(x,y)$,
and $\sigma_c$, $\sigma_m$ are the root-mean-squared structural and magnetic
roughnesses, respectively.
Note that the correlation term $U_{\nu\mu}$ due to the roughness in the 
reflection coefficient contains only independent contributions of
chemical and magnetic roughnesses expressed via $\sigma_c$ and $\sigma_m$,
respectively.
According to Eq. (\ref{eq:cross_section_dwba}), the diffuse scattering
must contain the cross-correlation component due to the term
$\Bigl<|{\cal T}^{fi}|^2 \Bigr>$.

A better approximation than Eq. (\ref{eq:RUV_mat}) may be obtained by using the
rough-interface reflection coefficient $R_{\nu\mu}$ instead of the
smooth-interface $R^{(0)}_{\nu\mu}$ in the wave functions of Eqs. (\ref{eq:E_ki}) and
(\ref{eq:E_kf}), thus getting a self-consistent matrix equation in
terms of the $2\times 2$ matrices, ${\bf R}$, ${\bf U}$, ${\bf V}$. 
This leads to
\bb\label{eq:RUV_eq}
{\bf R} = {\bf R}^{(0)} + {\bf U} + {\bf V}{\bf R},
\ee
whose solution is
\bb\label{eq:ref_final}
{\bf R} = ({\bf 1}-{\bf V})^{-1}( {\bf R}^{(0)} + {\bf U} ).
\ee
Similarly, for the reverse interface between upper resonant magnetic and
lower nonmagnetic layers, we can have the same solution as Eq. (\ref{eq:ref_final})
from Eqs. (\ref{eq:chi0_mat_rn}) and (\ref{eq:T_R_rn}).
The explicit expressions of ${\bf U}$, ${\bf V}$, ${\bf R}^{(0)}$ matrices 
in Eq. (\ref{eq:ref_final}) for both cases are given in Appendix C.

For nonmagnetic interfaces, the matrices are all diagonal($\sigma$ and $\pi$
polarizations are decoupled), and it has been shown that Eq. (\ref{eq:ref_final}) 
leads to the familiar Nevot-Croce form\cite{nevot} for the reflection coefficient, 
i.e.,
\bb\label{eq:nevot_croce}
R = R^{(0)} e^{-2|k_z||k_z^t|\sigma_c^2}.
\ee
The derivation of this is shown in Appendix D. 
For the magnetic interface, this simplified form for the reflection coefficient 
does not have any analogue.
Nevertheless, at sufficiently large values of $q_z$, the reflectivity takes the
familiar Gaussian form $R^{(0)} e^{-q_z^2\sigma_{\rm eff}^2}$. 
However, $\sigma_{\rm eff}^2$ does not always take the form predicted 
by the simple kinematical theory [i.e., $\sigma_c^2$ for $\sigma\rightarrow\sigma$
reflectivity, $\sigma_m^2$ for $\sigma\rightarrow\pi$ reflectivity, and
$\frac{1}{2}(\sigma_c^2 + \sigma_m^2 )$ for $(I_+ - I_-)$ in the case of
circularly polarized x-rays] as we shall see in the numerical example shown
below, which provides a counter-illustration of the rule that, at large $q_z$,
the DWBA becomes identical to the Born approximation or kinematical limit.

For circularly polarized incident x-rays with ${\bf\hat{e}}_{\pm}(\vec{k}_i) =
\left({\bf\hat{e}}_{\sigma}(\vec{k}_i)\pm i {\bf\hat{e}}_{\pi}(\vec{k}_i) 
\right)/\sqrt{2}$, the reflection amplitudes for $\sigma$- and $\pi$-polarization 
are given by
\bb\label{eq:ref_amplitude}
\left( \begin{array}{c}
R_{\sigma} \\ R_{\pi} \end{array} \right) =
{\bf R} \left( \begin{array}{c} \frac{1}{\sqrt{2}} \\ \pm \frac{i}{\sqrt{2}}
\end{array} \right),
\ee
where ${\bf R}$ is the $2\times 2$ matrix reflection coefficient 
in Eq. (\ref{eq:ref_final}).
The reflected intensities without polarization analysis for the outgoing beam,
$I=\sqrt{|R_{\sigma}|^2 +|R_{\pi}|^2}$, can be then evaluated for the opposite
helicities of incident beams  as
\bb\label{eq:difference_inten}
I_+ - I_- = 2~ {\rm Im} [R_{11}R^{\ast}_{12} + R_{21}R^{\ast}_{22}],
\ee
where $R_{ij}$ is the $ij$-element of the $2\times 2$ matrix ${\bf R}$.

Since Parratt's recursive formula for multiple interfaces includes only
reflection coefficients, its extension to the rough interface case does not
need the transmission coefficient to account for interface roughness. 
On the other hand, in our case where the fields are not scalars, the
transmission coefficients are requisite to calculate recursive 2$\times$2
matrix formulae for multiple magnetic interfaces, which will be discussed in
Sec. VII. 
For completeness, therefore, let us now calculate the transmission
coefficient $T_{j\mu}$ from a rough interface. 
In the spirit of Ref.~\onlinecite{boer94}, we assume for ${\bf E}({\bf k}_i, \mu)$ 
and ${\bf E}^T (-{\bf k}_f, j)$ the functional forms analytically continued both
for $z>0$ and for $z<0$ as follows:
\bb\label{eq:E_ki_trans}
{\bf E}({\bf k}_i, \mu) = \sum_{j^{\prime}=1,2}
                           T^{(0)}_{j^{\prime}\mu}({\bf k}_i)
{\bf\hat{e}}_{j^{\prime}} e^{i{\bf k}_i^t(j^{\prime})\cdot{\bf r}},\\
\label{eq:E_kf_trans}
{\bf E}^T(-{\bf k}_f, j) =
\sum_{\nu=\sigma,\pi} T^{(0)\ast}_{\nu j}(-{\bf k}_f) {\bf\hat{e}}_{\nu}
                                          e^{-i{\bf k}_f^{\ast}\cdot{\bf r}},
\ee
where $T^{(0)\ast}_{\nu j}(-{\bf k}_f)$ in Eq. (\ref{eq:E_kf_trans}) denotes the
transmission coefficient ``from'' a magnetic(anisotropic) medium ``to'' a
nonmagnetic (isotropic) one, whose explicit form is given in Appendix A. 
For the smooth surface, 
the scattering matrix element between the eigenstates $|-{\bf k}_f^T , j>$ and
$|{\bf k}_i, \mu>$ can be then written as
\bb\label{eq:chi0_mat_trans}
k_0^2<-{\bf k}^T_f, j | \bm{\chi}^{(0)} | {\bf k}_i, \mu>
&=& i{\cal A} k_0^2\delta_{k_{ix}k_{fx}} \delta_{k_{iy}k_{fy}} \nonumber \\
&\times& \sum_{\nu} T^{(0)}_{\nu j}(-{\bf k}_f)
     \sum_{j^{\prime}} T^{(0)}_{j^{\prime}\mu}({\bf k}_i)
        \sum_{\alpha\beta} e^{\ast}_{\nu\alpha}
                (\chi_1 \delta_{\alpha\beta} + \chi_{\alpha\beta}^{(2)})
                 e_{j^{\prime}\beta} \nonumber \\
&\times& \int^{0}_{-\infty} dz e^{-i (- k_{fz}-k^t_{iz}(j^{\prime}))z },
                                                           \nonumber \\
&=& 4i {\cal A} k^t_{iz}(j) T^{(0)}_{j\mu}({\bf k}_i) \delta_{k_{ix}k_{fx}}
                                               \delta_{k_{iy}k_{fy}},
\ee
where $T^{(0)}_{j\mu}({\bf k}_i)$ is the transmission coefficient for the smooth
surface, as defined in Eq. (\ref{eq:E_ki}).
The details of Eqs. (\ref{eq:chi0_mat_trans}) are given in Appendix B.

In comparison with Eq. (\ref{eq:chi0_mat_trans}) for the smooth surface, 
the scattering matrix element for the rough surface, as shown in
Eq. (\ref{eq:scatt_mat_element}), can be analogously defined by
\bb \label{eq:T_T}
<{\bf k}_f, j| {\cal T}|{\bf k}_i, \mu> 
	= 4 i {\cal A} k_{iz}^t (j) T_{j\mu}({\bf k}_i) 
		\delta_{k_{ix}k_{fx}}\delta_{k_{iy}k_{fy}},
\ee                                                                    
where $T_{j\mu}({\bf k}_i)$ denotes the transmission coefficient for the 
rough surface. 

For the statistically averaged amplitude $\Bigl< {\cal T}^{fi}\Bigr>$,
we obtain
\bb\label{eq:chi_cm_mat_tran}
\Bigl< k_0^2 <-{\bf k}_f^T, j|{\bf\Delta}^{c,m}|{\bf k}_i, \mu > \Bigr> 
&=& i {\cal A} k_0^2 \sum_{\nu} T^{(0)}_{\nu j}(-{\bf k}_f)  \nonumber \\ 
&\times& \sum_{j^{\prime}} T^{(0)}_{j^{\prime}\mu}({\bf k}_i) \sum_{\alpha\beta}
\frac{e_{\nu\alpha}^{\ast} \Delta^{c,m}_{\alpha\beta}e_{j^{\prime}\beta}}
		{q_{3z}(j^{\prime})}
                \biggl[\Bigl<e^{-i q_{3z}(j^{\prime})\delta z_{c,m}(x,y)}\Bigr> -1 \biggr],
\ee
and
\bb\label{eq:qz_3}
q_{3z}(j^{\prime}) = - k_{fz} - k_{iz}^t(j^{\prime}).
\ee

From Eqs. (\ref{eq:chi0_mat_trans})-(\ref{eq:T_T}) and (\ref{eq:chi_cm_mat_tran}),
we see that we can write the scattering matrix element in the DWBA, 
as shown in Eq. (\ref{eq:scatt_mat_element}), as
\bb\label{eq:TV_mat}
T_{j\mu} = T^{(0)}_{j\mu} + \sum_{j^{\prime}=1,2} V^{\prime}_{jj^{\prime}}
T^{(0)}_{j^{\prime}\mu},
\ee
where
\bb\label{eq:V_mat}
V^{\prime}_{jj^{\prime}} &=& \sum_{\nu}\frac{T^{(0)}_{\nu j}(-{\bf k}_f)}
	{4 k^t_{iz}(j)} \frac{k^2_0}{q_{3z}(j^{\prime})}
\biggl[ (\chi_1-\chi_0)\sum_{\alpha}
         e^{\ast}_{\nu\alpha} e_{j^{\prime}\alpha}
         \Bigl[ e^{-\frac{1}{2}q_{3z}^2(j^{\prime})\sigma_c^2} - 1 \Bigr]
                                                         \nonumber \\
  &~&~~~~~~~~+ \sum_{\alpha\beta}e^{\ast}_{\nu\alpha}\chi^{(2)}_{\alpha\beta}
                        e_{j^{\prime}\beta}
        \Bigl[ e^{-\frac{1}{2}q_{3z}^2(j^{\prime})\sigma_m^2} - 1 \Bigr]
	\biggr].
\ee

In the same way as we did for the reflection coefficient, 
using the rough-interface transmission coefficient $T_{j\mu}$
instead of the smooth-interface $T_{j\mu}^{(0)}$ in the right side of 
Eq. (\ref{eq:TV_mat}), thus getting a self-consistent matrix equation 
in terms of the $2\times 2$ matrices, ${\bf T}$, ${\bf V^{\prime}}$, gives
\bb\label{eq:TV_eq}
{\bf T} = {\bf T}^{(0)} + {\bf V^{\prime}}{\bf T},
\ee
whose solution is
\bb\label{eq:trans_final}
{\bf T} = ({\bf 1}-{\bf V^{\prime}})^{-1} {\bf T}^{(0)}.
\ee
Similarly, for the reverse interface between upper resonant magnetic and 
lower nonmagnetic layers, we can also have the same solution as 
Eq. (\ref{eq:trans_final}).
The explicit expressions of ${\bf V^{\prime}}$ and ${\bf T}^{(0)}$
matrices in Eq. (\ref{eq:trans_final}) for both cases are given 
in Appendix C.

For nonmagnetic interfaces, it is shown in Appendix D that 
Eq. (\ref{eq:trans_final}) reduces to
\bb\label{eq:vidal_vincent}
T = T^{(0)} e^{\frac{1}{2}\left( |k_z|-|k_z^t|\right)^2\sigma_c^2},
\ee
which has been found by Vidal and Vincent.\cite{vidal}


\section{Numerical Examples for A Single Magnetic Surface}

We now illustrate numerical examples of the above formulae calculated for a Gd
surface with varying degrees of structural and magnetic roughness. 
We have considered only the case where the magnetization vector is aligned 
along the sample surface in the scattering plane 
in order to enhance the magnetic effect.

\begin{figure}
\epsfxsize=12cm
\centerline{\epsffile{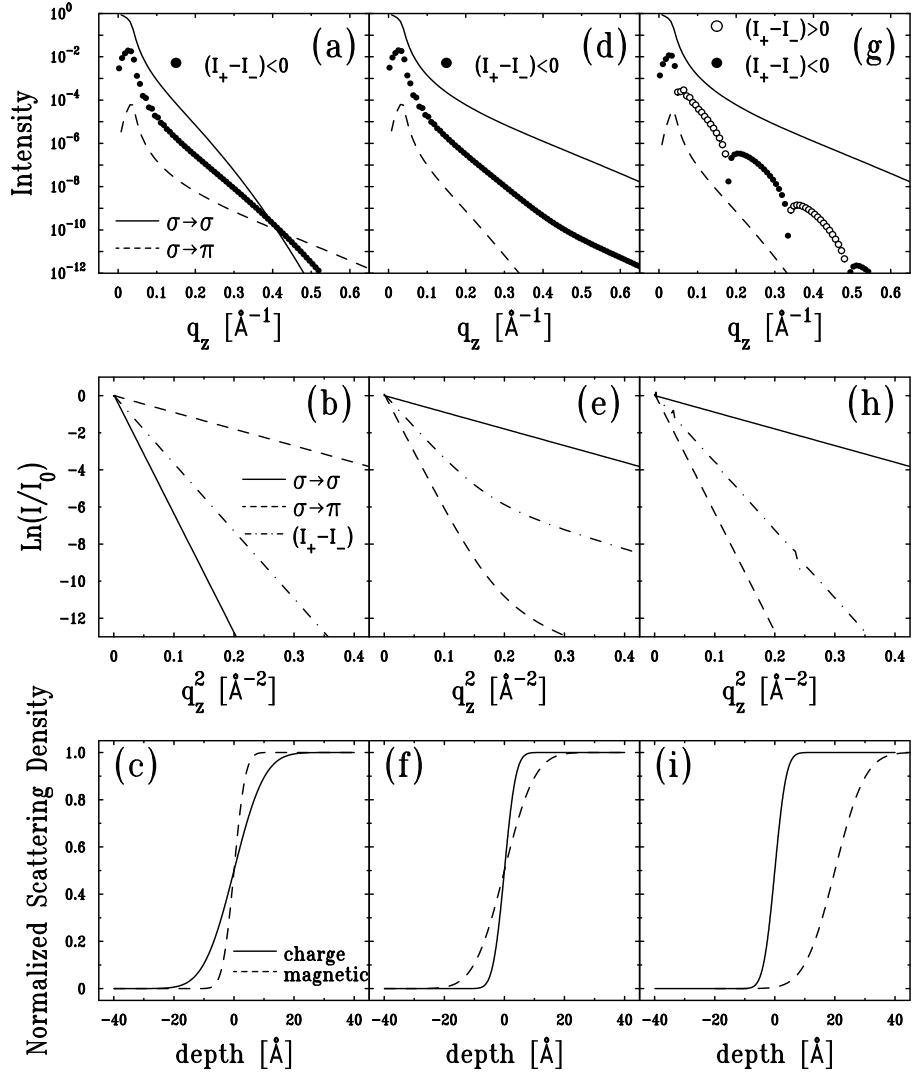}}
\caption{\label{fig-surf-refl}
Calculated x-ray resonant magnetic reflectivities at the Gd L$_3$-edge (7243
eV) from Gd surfaces with different interfacial widths for structural
($\sigma_c$) and magnetic ($\sigma_m$) interfaces:
(a)-(c) $\sigma_c = 8$\AA,~$\sigma_m = 3$\AA.~
(d)-(f) $\sigma_c = 3$\AA,~$\sigma_m = 8$\AA.~
(g)-(i) same as (d)-(f), but with a 20\AA~magnetically dead layer.
Top panel: reflected intensities of the $\sigma\rightarrow\sigma$ (solid
lines) and $\sigma\rightarrow\pi$ (dashed lines) channels, and the
differences between the reflected intensities for right- ($I_+$) and left-
($I_-$) circularly polarized incident beams (circles). Middle panel:
Natural logarithms of the reflectivities with interface roughnesses
normalized to those from ideal systems without roughness as a function of the
square of the wave-vector transfer. Solid, dashed, and dot-dashed lines
represent $\sigma\rightarrow\sigma$ and $\sigma\rightarrow\pi$ scattering,
and the differences between $I_+$ and $I_-$, respectively. Bottom panel:
Normalized scattering density profiles for charge (solid lines) and magnetic
(dashed lines) scattering.}
\end{figure}

Figure \ref{fig-surf-refl} shows the x-ray resonant magnetic reflectivities
calculated at the Gd L$_3$-edge (7243 eV) from Gd surfaces with different
interfacial widths for structural ($\sigma_c$) and magnetic ($\sigma_m$) interfaces. 
In Fig. \ref{fig-surf-refl}(a)-(c), the interfacial width of the
structural interface is larger than that of the magnetic interface, that is,
$\sigma_c = 8$ \AA~and $\sigma_m = 3$ \AA.~
On the other hand, in Fig.
\ref{fig-surf-refl}(d)-(f), the interfacial widths are reversed, that is,
$\sigma_c = 3$ \AA~and $\sigma_m = 8$ \AA.~ 
In the kinematical approximation(BA)
$\sigma\rightarrow\sigma$ scattering (solid lines in the top panels of Fig.
\ref{fig-surf-refl}) corresponds to pure charge scattering, and
$\sigma\rightarrow\pi$ scattering (dashed lines in the top panels of Fig.
\ref{fig-surf-refl}) to pure magnetic scattering, and the differences between
the reflected intensities for right- ($I_+$) and left- ($I_-$) circularly
polarized incident beams (circles in Fig. \ref{fig-surf-refl})
correspond to the interferences between charge and magnetic scattering.

Kinematically, the reflected intensities from each scattering channels are
proportional to a simple Gaussian form, $\exp(-\sigma^2 q_z^2)$, where
$\sigma$ is the interfacial width of corresponding scattering channel, i.e.,
$\sigma_c$ for $I_{\sigma\rightarrow\sigma}$, $\sigma_m$ for
$I_{\sigma\rightarrow\pi}$, and $\sqrt{(\sigma_c^2+\sigma_m^2)/2}$ for $(I_+-I_-)$. 
The middle panel of Fig. \ref{fig-surf-refl} shows natural
logarithms of the reflectivities from rough interfaces normalized to those
from ideal systems without roughness as a function of the square of the
wave vector, $q_z^2$, whose slopes are then equal to the squares of the
interfacial widths for their corresponding scattering channels. 
In Fig. \ref{fig-surf-refl}(b), the slopes obtained from our dynamical calculation 
for the case of $\sigma_c = 8$ \AA~and $\sigma_m = 3$ \AA~ show good agreement
with the kinematical results mentioned above. 
On the other hand, in Fig. \ref{fig-surf-refl}(e), 
the slopes of $I_{\sigma\rightarrow\pi}$ and $(I_+-I_-)$ for the opposite case, 
$\sigma_c = 3$ \AA~and $\sigma_m = 8$ \AA,~ are
not equal to the squares of their corresponding interfacial widths but follow
the slope of $I_{\sigma\rightarrow\sigma}$ at high $q_z$'s.

This indicates that the kinematical argument mentioned above, i.e.,
one-to-one correspondence such as $\sigma\rightarrow\pi$ channel to pure
magnetic scattering, is no longer valid for such a case of larger magnetic
interfacial width, as shown in Fig. \ref{fig-surf-refl}(e). 
In other words, both contributions from charge and magnetic scattering 
should be taken into account for every scattering channel, 
which is naturally included in the dynamical theory (such as 
our self-consistent method). 
In the case shown in Fig. \ref{fig-surf-refl}(e), 
since the charge-scattering channel 
is much stronger than the magnetic-scattering channel 
and also drops off much more slowly with $q_z$ due to decreased roughness, 
there is conversion of $\sigma\rightarrow\pi$ polarization 
at larger $q_z$ even when the ``pure'' magnetic scattering has become negligible 
in the kinematical limit, 
because of magnetic scattering out of the still strong charge channel. 
Thus the $\sigma\rightarrow\pi$ and $(I_+ - I_-)$ reflections will asymptotically
decay at a rate governed by the decay of the charge channel, which is
determined by $\sigma_c$ alone.

However, it is not easy to find a physical system where a magnetic
interfacial width is larger than the structural one at the same interface, as
shown in Fig. \ref{fig-surf-refl}(f). 
Instead, such a rougher magnetic interface can occur in a magnetic system, 
where a magnetically ``dead'' layer exists near the top surface 
and so the average position of the magnetic interface may not coincide 
with that of the structural interface, as shown in Fig. \ref{fig-surf-refl}(i). 
In Fig. \ref{fig-surf-refl}(g) $(I_+-I_-)$ (circles) shows 
an oscillation due to a magnetically dead layer with its thickness of 20 \AA.~ 
In this case, the slopes in Fig. \ref{fig-surf-refl}(h) follow again 
the kinematical result mentioned above because the magnetic interface and 
the structural one are separated spatially.

As a further check on our calculations, we have calculated the reflectivity
by dividing the error-function profile, 
as shown in the bottom panel of Fig. \ref{fig-surf-refl}, 
into many very thin slices and using the 2$\times$2 recursive matrix formulae 
without any roughness assumptions.\cite{stepanov}
We found that the results using this slice method are exactly the same as
those from our self-consistent method assuming Gaussian height distributions
in Fig. \ref{fig-surf-refl}. 
Thus our self-consistent method based on the DWBA produces very accurate results
for the x-ray resonant magnetic reflectivity and much faster computationally.


\section{Multiple Magnetic Interfaces}

For a multilayer with multiple interfaces, each layer can be characterized by
its dielectric susceptibility tensor $\chi_{\alpha\beta, n}$ for the $n$-th
layer, which can be $\chi_{\alpha\beta, n}=\chi_n \delta_{\alpha\beta}$ for
nonmagnetic (isotropic) layers and 
$\chi_{\alpha\beta, n}=\chi_n\delta_{\alpha\beta} + \chi^{(2)}_{\alpha\beta, n}$ 
for magnetic (anisotropic) layers. 
For each rough interface, we can use the self-consistent DWBA to define 
the reflection and transmission coefficients, in the same way as in Sec. V, 
which are given by
\bb\label{eq:RT_M_rough}
{\bf R}_n &=& ({\bf I}-{\bf V}_n)^{-1}({\bf R}^{(0)}_n +{\bf U}_n) = \tilde{M}^{rt}_n,
		                                         \nonumber \\
{\bf T}_n &=&({\bf I}-{\bf V^{\prime}}_n)^{-1}{\bf T}^{(0)}_n = \tilde{M}^{tt}_n,
\ee
where ${\bf R}_n$, ${\bf T}_n$ are the reflection and transmission coefficients
for the $n$-th rough interface, and ${\bf R}^{(0)}_n$, ${\bf T}^{(0)}_n$ are those for 
the corresponding smooth interface. 
The explicit expressions for ${\bf R}^{(0)}_n$, ${\bf T}^{(0)}_n$, ${\bf U}_n$,
${\bf V}_n$, and ${\bf V^{\prime}}_n$ matrices in Eq. (\ref{eq:RT_M_rough}) 
are given in Appendix C,
depending on whether the upper and lower layers on the $n$-th interface are 
nonmagnetic or magnetic layers, respectively.

By analogy with the recursion relation for the coupled waves derived for the
smooth interfaces in Appendix E (originally developed 
by Stepanov and Sinha\cite{stepanov}),
introducing $\tilde{W}^{pq}$ matrices for the rough interfaces,
we may derive the recursion relation  
analogous to Eq. (\ref{eq:WM_mat_ml}), obtaining
\bb\label{eq:W_M_rough}
\tilde{W}^{tt}_{n+1} &=& \tilde{A}_n \tilde{W}^{tt}_n, \nonumber \\
\tilde{W}^{tr}_{n+1} &=& \tilde{M}^{tr}_{n+1} + \tilde{A}_n \tilde{W}^{tr}_n
                                   \tilde{M}^{rr}_{n+1}, \nonumber \\
\tilde{W}^{rt}_{n+1} &=& \tilde{W}^{rt}_n + \tilde{B}_n \tilde{M}^{rt}_{n+1}
                                       \tilde{W}^{tt}_n, \nonumber \\
\tilde{W}^{rr}_{n+1} &=& \tilde{B}_n \tilde{M}^{rr}_{n+1},
\ee
where $\tilde{A}_n$ and $\tilde{B}_n$ are defined by
\bb\label{eq:AB_rough}
\tilde{A}_n = \tilde{M}^{tt}_{n+1} \left( 1 - \tilde{W}^{rt}_n
                         \tilde{M}^{rt}_{n+1}\right)^{-1}, \nonumber \\
\tilde{B}_n = \tilde{W}^{rr}_n \left( 1 - \tilde{M}^{rt}_{n+1}
                               \tilde{W}^{tr}_n\right)^{-1}.
\ee

Finally, the specular reflectivity of a magnetic multilayer with rough
interfaces can be obtained by
\bb\label{eq:ref_maplitude_ml}
R_0 = \tilde{W}^{rt}_N T_0.
\ee
To calculate the sum and difference in the reflectivities for $(+)$ and $(-)$
circularly polarized incident x-rays, substituting $T_0 =
\frac{1}{\sqrt{2}}(1,\pm i)$ in a similar way to 
Eqs. (\ref{eq:ref_amplitude}) and (\ref{eq:difference_inten}) yields
\bb\label{eq:sum_diff_inten}
I_+ + I_- &=&  |(\tilde{W}^{rt}_N)_{11}|^2 + |(\tilde{W}^{rt}_N)_{12}|^2
+ |(\tilde{W}^{rt}_N)_{21}|^2 +|(\tilde{W}^{rt}_N)_{22}|^2, \nonumber \\
I_+ - I_- &=& 2~ {\rm Im}\left[ (\tilde{W}^{rt}_N)_{11}
  (\tilde{W}^{rt}_N)^{\ast}_{12}
+ (\tilde{W}^{rt}_N)_{21} (\tilde{W}^{rt}_N)^{\ast}_{22} \right],
\ee
where $(\tilde{W}^{rt}_N)_{ij}$ is the $ij$-element of the $2\times 2$ matrix
$\tilde{W}^{rt}_N$.

The above suggested approach to calculating the effects of roughness
in multilayers on specular reflectivity is an approximation analogous
to those used previously in several publications on 
charge-only roughness.\cite{holy94,boer91,press93,stepanov94}
Basically, it corresponds to averaging the reflection coefficient (or the
scattering matrix) of each interface over the interface roughness.
The comparison with the results of rigorous ``slicing method'' made
in Ref. \onlinecite{stepanov94} has proven that 
such an approximation works very well. 
A possible reason for the excellent validity of this approximation is
that the roughness effect is mainly displayed at greater incidence
angles, where the reflection is small and the multiple scattering
can be neglected (the total reflection amplitude is a linear sum of
contributions from individual interfaces).
Note that, since we are considering the coherent scattering which
involves only the statistical average of 
the scattering amplitude in Eq. (\ref{eq:cross_section_dwba}),
there is no contribution from any cross-interface
correlations of roughness. 
This will not be the case with diffuse (off-specular) scattering.\cite{paperII}


\section{Numerical examples for multiple interfaces}

We present here numerical examples for x-ray resonant magnetic reflectivity
from a Gd/Fe multilayer using the above formulae. 
Since Gd/Fe multilayers (MLs) have vastly different Curie temperatures 
and strong interfacial coupling of Gd and Fe, 
these systems give rise to complex magnetic structures depending on the layer 
thickness, temperature, and applied magnetic field.\cite{camley}
Due to the advantage of Gd L$-$edge resonances available in the hard x-ray
regime, several experimental studies from these Gd/Fe MLs have been performed
using x-ray resonant magnetic reflectivity 
measurements.\cite{xrms-drlee,ishimatsu99,haskel01} 
Again, we have considered
only the case where the magnetization vector ${\bf M}\parallel {\bf\hat{x}}$.

\begin{figure}
\epsfxsize=10cm
\centerline{\epsffile{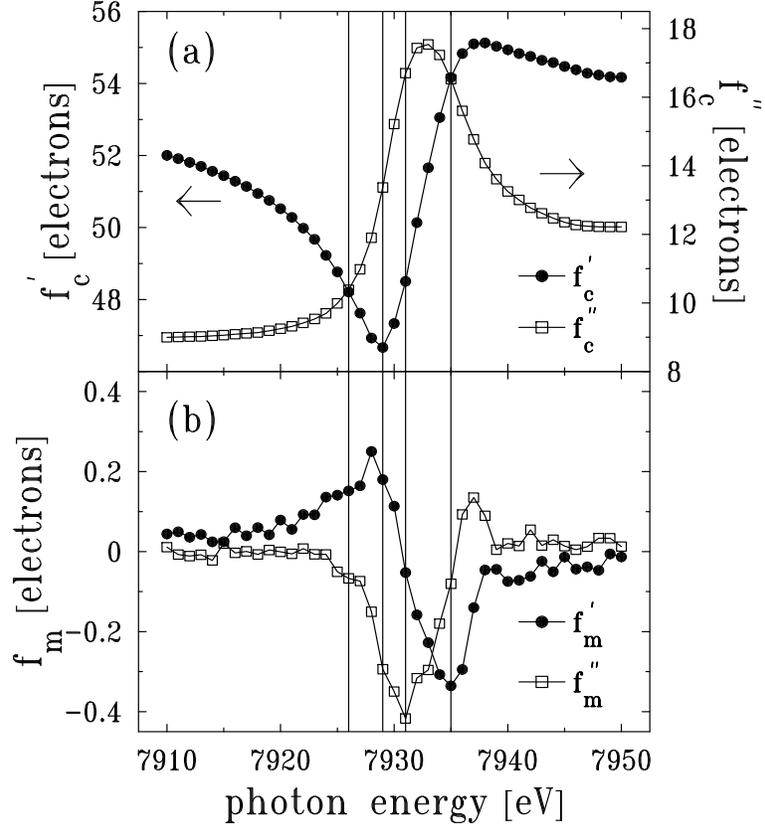}}
\caption{ \label{fig-fac}
Charge (a) and magnetic (b) x-ray scattering amplitudes, $f_{c,m}$ around the
Gd L$_2$-edge obtained from the absorption measurements for a [Gd(51
\AA)/Fe(34 \AA)]$_{15}$ multilayer. The vertical lines indicate the photon
energies, where the x-ray resonant magnetic reflectivities in Fig.
\ref{fig-energy} were calculated. }
\end{figure}

We have used the experimentally determined values for charge and magnetic
resonant scattering amplitudes, $f_{c,m} = f^{\prime}_{c,m} + i
f^{\prime\prime}_{c,m}$, at the resonant energy. 
The energy dependence of the absorption coefficient for opposite helicites, 
$\mu^{\pm}(E)$, were measured from a [Gd(51 \AA)/Fe(34 \AA)]$_{15}$ multilayer, 
which will be discussed below as an experimental example. 
The edge-step normalized $f^{\prime\prime}_{c,m}$ were obtained 
from the charge and magnetic absorption coefficients, 
$\mu_{c,m}$ [$\mu_c = (\mu^+ + \mu^-)/2$, $\mu_m =\mu^+ - \mu^-$], 
through the optical theorem, $f^{\prime\prime}_{c,m}\propto \mu_{c,m}$. 
Their absolute values were determined using the tabulated
bare-atom scattering amplitudes away from resonance. 
Real parts were obtained from differential Kramers-Kronig transforms of 
imaginary parts. 
Figure \ref{fig-fac}(a) and (b) show the charge and magnetic scattering amplitudes
around the Gd L$_2$-edge obtained in such absorption measurements. 
These values are in good agreement with the calculated ones from the listed values
of $A$ and $B$ in Eq. (\ref{eq:ABC}) obtained from Ref.~\onlinecite{hamrick}. 
For consistency of the definitions, it should be mentioned that
the $f^{\prime\prime}_{c,m}$ used here correspond to 
Im[$A$, $B$] in Eq. (\ref{eq:ABC}),
whereas the $f^{\prime}_{c,m}$ correspond to $-$Re[$A$, $B$], respectively.

\begin{figure}
\epsfxsize=14cm
\centerline{\epsffile{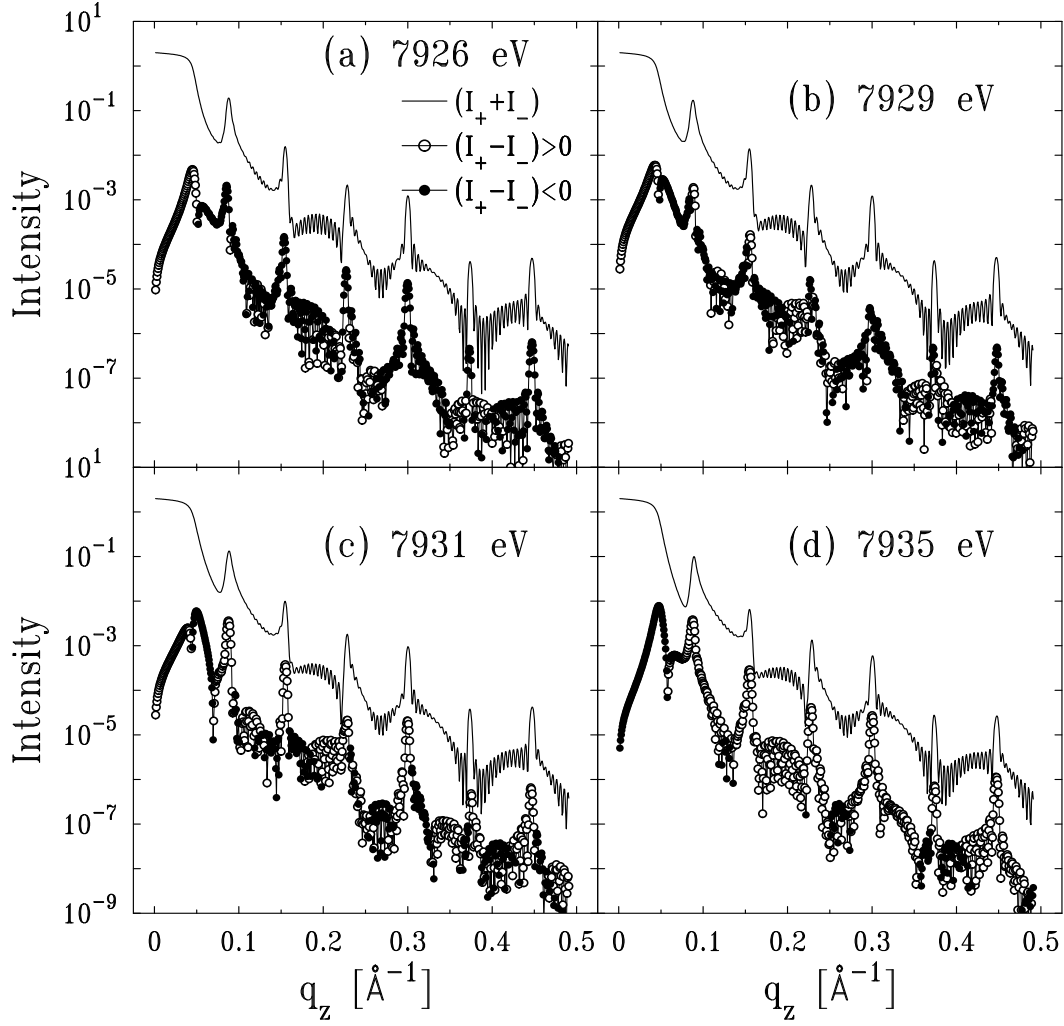}}
\caption{ \label{fig-energy}
Calculated x-ray resonant magnetic reflectivities from a
[Gd(51\AA)/Fe(34\AA)]$_{15}$ multilayer for different incident photon
energies indicated in Fig. \ref{fig-fac}: (a) 7926 eV, (b) 7929 eV, (c) 7931
eV, and (d) 7935 eV. Both structural (charge) and magnetic interface
roughnesses are $\sigma_{c,m}= 4.7$ \AA~and 3.6 \AA~for Fe/Gd and Gd/Fe
interfaces, respectively. The solid lines represent $(I_++I_-)$ intensities
and open (filled) circles represent the positive (negative) values of
$(I_+-I_-)$ intensities.}
\end{figure}

Figure \ref{fig-energy} shows the calculated x-ray resonant magnetic
reflectivities from a [Gd(51 \AA)/Fe(34 \AA)]$_{15}$ multilayer for different
incident x-ray energies indicated in Fig. \ref{fig-fac}: (a) 7926 eV, (b)
7929 eV, (c) 7931 eV, and (d) 7935 eV. 
The lines and symbols represent the sum and difference in the reflected intensities 
for $(+)$ and $(-)$ circularly polarized incident x-rays, respectively, 
calculated using Eq. (\ref{eq:sum_diff_inten}). 
Since the Gd/Fe multilayer was assumed to be sandwiched between Nb
buffer (100 \AA) and cap (30 \AA) layers, the Kiessig fringes between the
multilayer peaks in $(I_++I_-)$ intensities result from the interference of
the scattering of Nb layers and thus show little energy dependence around the
Gd absorption edge. 
On the other hand, $(I_+-I_-)$ intensities around the multilayer peaks show 
a clear energy dependence in signs and magnitudes relative to $(I_++I_-)$ intensities.
In Fig. 5(a) and (d) at which energies $f_m^{\prime\prime}$ becomes 
much smaller than $f_m^{\prime}$, the signs and relative magnitudes of 
$(I_+-I_-)$ intensities follow simply the energy
dependence of $f_m^{\prime}$ in Fig. \ref{fig-fac}(b), 
as expected in the kinematical approximation.\cite{osgood} 
At the energies close to the absorption edge where $f_m^{\prime\prime}$ cannot 
be neglected, however, one can hardly expect the signs and magnitudes of 
$(I_+-I_-)$ intensities to be obtained directly from the values of $f_m^{\prime}$ and 
$f_m^{\prime\prime}$ in Fig. \ref{fig-fac}(b). 
Therefore, quantitative analysis on x-ray resonant magnetic reflectivity data 
at the resonant energy requires accurate calculation taking
into account refraction and multiple scattering effects using dynamical
theory, such as our self-consistent method presented above.

\begin{figure}
\epsfxsize=14cm
\centerline{\epsffile{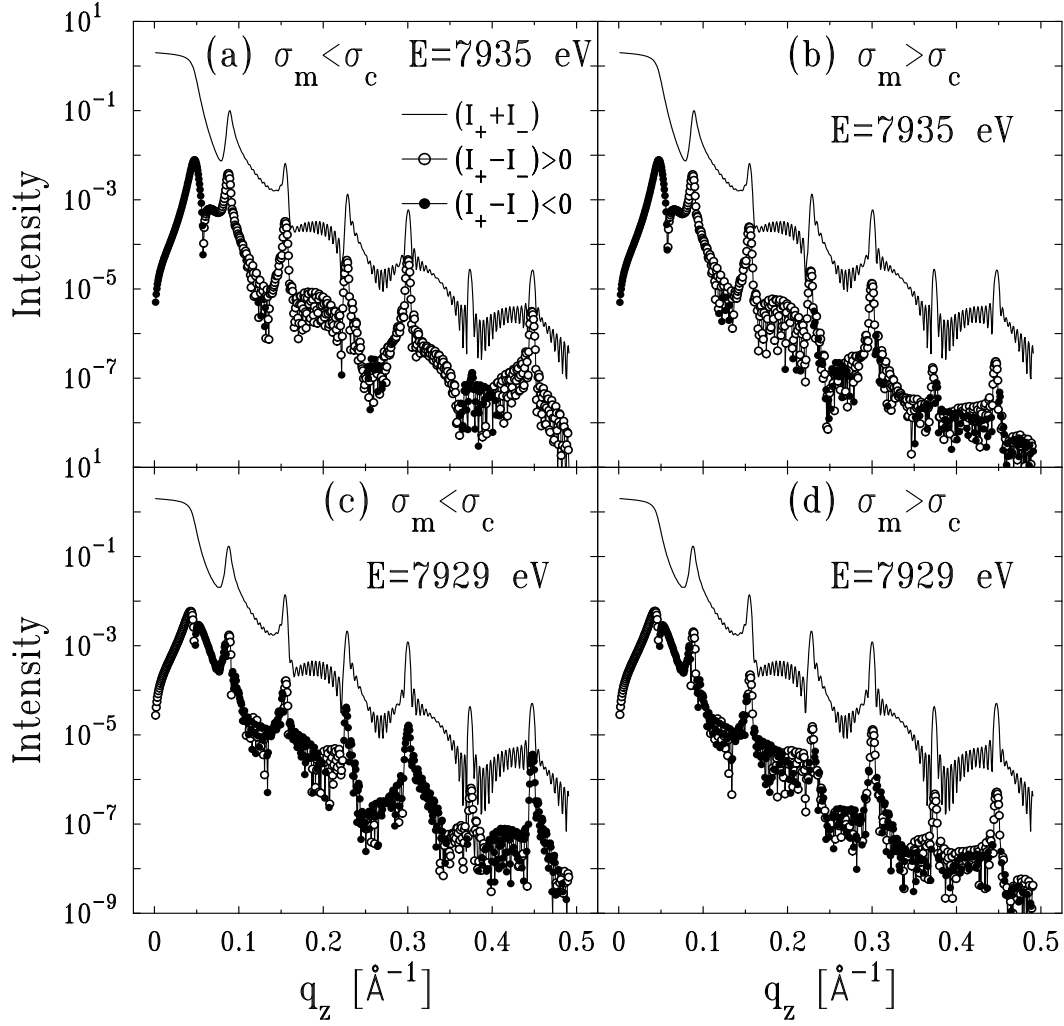}}
\caption{ \label{fig-rms}
Calculations with different magnetic interface roughnesses: (a) and (c)
$\sigma_m$ = 2.1 \AA~, and (b) and (d) $\sigma_m$ = 6.2\AA. All other
parameters and symbols are same as those in Fig. \ref{fig-energy}. }
\end{figure}

In order to study the effect of the magnetic roughness amplitude, $(I_+-I_-)$
intensities for two cases, $\sigma_m < \sigma_c$ and $\sigma_m > \sigma_c$,
have been calculated, as shown in Fig. \ref{fig-rms}. 
The calculations for $\sigma_m = \sigma_c$ have been shown in Fig. \ref{fig-energy}. 
For all cases, the charge roughness amplitudes were assumed to be 
$\sigma_{c,{\rm Fe/Gd}} = 4.7$ \AA and $\sigma_{c,{\rm Gd/Fe}} = 3.6$ \AA. 
At the energy of 7935 eV, the intensities of $(I_+-I_-)$ around the multilayer peaks 
are proportional to a simple Gaussian form, exp($-\sigma^2 q_z^2$), as shown in
Figs. \ref{fig-energy}(d), \ref{fig-rms}(a), and \ref{fig-rms}(b). 
This is consistent with the kinematical calculations,\cite{osgood} 
and $\sigma$ for $(I_+-I_-)$ corresponds to $\sqrt{(\sigma_c^2+\sigma_m^2)/2}$ 
as given by the kinematical argument. 
On the other hand, at the energy of 7929 eV where $f_m^{\prime\prime}$ cannot 
be neglected, such a kinematical argument is no longer valid. 
Comparing Figs. \ref{fig-energy}(b), \ref{fig-rms}(c), and
\ref{fig-rms}(d), we can see that the magnitudes of $(I_+-I_-)$ peak
intensities do not follow a Gaussian form, exp($-\sigma^2 q_z^2$), but their
signs change from negative (filled circles) to positive (open circles) values. 
This indicates that $(I_+-I_-)$, which is known to be the
charge-magnetic interference scattering in the kinematical
theory,\cite{blume88} is sensitive even to the interference between charge
and magnetic roughness amplitudes. 
However, it should be mentioned again that
this result cannot be reproduced by the kinematical calculation but only by
the dynamical one presented above.

\begin{figure}
\epsfxsize=12cm
\centerline{\epsffile{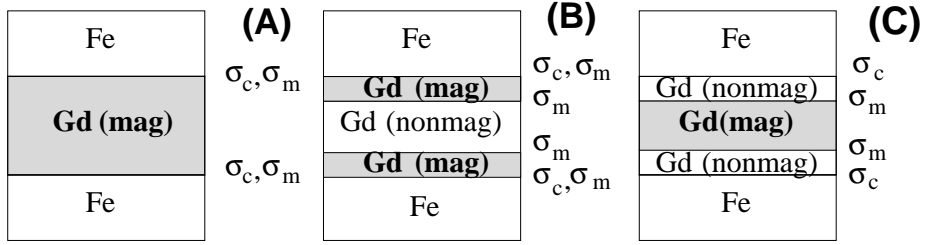}}
\caption{ \label{mag-geo}
Models of magnetic structures in Gd layers. Uniform magnetization (A),
ferromagnetic moments only near the Gd/Fe interfaces (B), and ferromagnetic
moments near the centers of Gd layers between magnetically dead layers (C).
While interfaces with ``$\sigma_c$, $\sigma_m$'' represent both structurally
and magnetically mixed interfaces, interfaces with ``$\sigma_c$'' (or
``$\sigma_m$'') represent purely structural (or magnetic) interfaces.}
\end{figure}

Let us now consider the case where the magnetic structure in the resonant
layers may not coincide with the chemical structure. 
For example, the ferromagnetic moments in Gd layers near Gd/Fe interfaces 
can be induced by the adjacent ferromagnetic Fe layers above the Curie temperature 
of Gd atoms,\cite{ishimatsu99,haskel01} or a magnetically ``dead layer'' may exist
at an interface between a ferromagnetic layer and an antiferromagnetic layer.
Here we assume simply three different magnetization depth profiles in the Gd
layers of a Gd/Fe multilayer, as shown in Fig. \ref{mag-geo}:  uniform
magnetization (A), 
ferromagnetic moments only near the Gd/Fe interfaces (B),
ferromagnetic moments near the centers of Gd layers between magnetically dead
layers (C).

\begin{figure}
\epsfxsize=14cm
\centerline{\epsffile{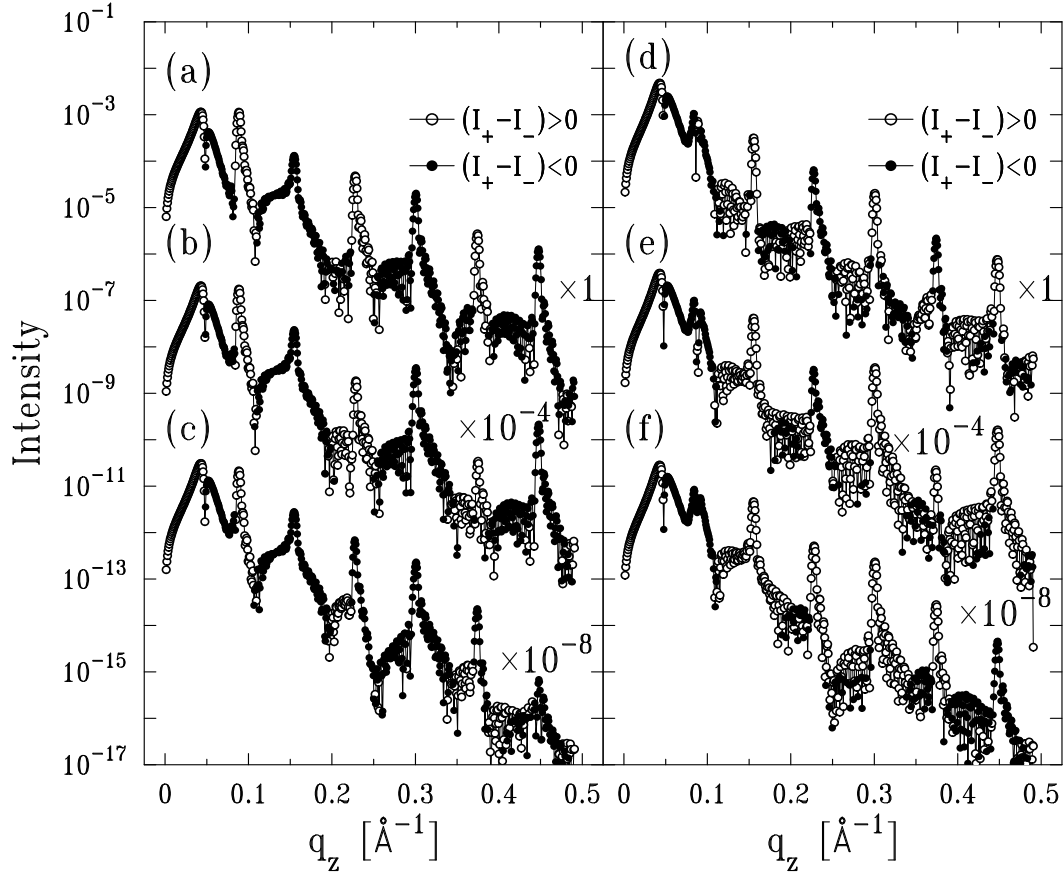}}
\caption{ \label{fig-layer}
Calculated $(I_+-I_-)$ intensities for different magnetization depth profiles
in Gd layers. In (a)-(c) ferromagnetic layers exist only near the Gd/Fe
interfaces [Fig. \ref{mag-geo}(B)], and their layer thicknesses are 4.6
\AA~(a), 8.4 \AA~(b), and 12.8 \AA~(c). In (d)-(f) ferromagnetic layers exist
in the middle of Gd layers and are sandwiched between magnetically dead layers
[Fig. \ref{mag-geo}(C)], and the layer thicknesses of the dead layers are 4.6
\AA~(d), 8.4 \AA~(e), and 12.8 \AA~(f). All magnetic roughness amplitudes are
$\sigma_m = 4.2$ \AA, which is effectively same as $\sigma_c$, and the photon
energy is $E=7929$ eV. All other parameters and symbols are same as those in
Fig. \ref{fig-energy}.}
\end{figure}

Figure \ref{fig-layer} shows the results of calculations of x-ray resonant
magnetic reflectivities from [Gd(51 \AA)/Fe(34 \AA)]$_{15}$ MLs with the
different magnetic structures of Fig. \ref{mag-geo}. 
We assumed all magnetic roughness amplitudes of 
$\sigma_m = 4.2$ \AA~(effectively same as $\sigma_c$)
and the photon energy of $E=7929$ eV. 
In Figs. \ref{fig-layer}(a)-(c), Gd layers were assumed to be magnetized 
only near the Gd/Fe interfaces [model (B)], 
and the thickness of each magnetized layer was assumed to be 4.6 \AA~(a), 
8.4 \AA~(b), and 12.8 \AA~(c). 
On the other hand, in Fig. \ref{fig-layer}(d)-(f), 
Gd layers were assumed to be magnetized in the middle of each Gd layer 
and sandwiched between magnetically dead layers [model (C)],
and the thickness of each dead layer was assumed to be 4.6 \AA~(d), 
8.4 \AA~(e), and 12.8 \AA~(f).

Unlike the case of uniform magnetization [model (A) in Fig. \ref{mag-geo}]
shown in Fig. \ref{fig-energy}(b), $(I_+-I_-)$ intensities in Fig.
\ref{fig-layer} for models (B) and (C) show no suppression in peak
intensities due to the charge-magnetic interference, as discussed above. 
This may be ascribed to a spatial separation between the charge and magnetic
interfaces in models (B) and (C), as shown in Fig. \ref{mag-geo}.

In addition, the signs and relative magnitudes of $(I_+ - I_-)$ intensities
at the multilayer peaks change remarkably as the thicknesses of magnetized
layers change. 
In general, the peak intensities of the $(m+n)-$th order ML
peak and its multiple orders are weak compared to other peak intensities when
the thickness ratio between two constituent layers is $n/m$. 
For example, in our Gd(51 \AA)/Fe(34 \AA) multilayer, the fifth peak 
corresponds to such a suppressed peak. 
Therefore, different thicknesses of magnetic layers readily change
the order of the suppressed peak in $(I_+ - I_-)$ intensities, as
shown in Fig. \ref{fig-layer}. 
On the other hand, the signs of $(I_+ - I_-)$ intensities for models 
(B) [Fig. \ref{fig-layer}(a)-(c)] and (C) [Fig. \ref{fig-layer}(d)-(f)] 
are opposite each other, because their magnetic
structures are exactly reversed.


\section{Experiments}

X-ray resonant magnetic reflectivities were measured from an 
Fe(34 \AA)/[Gd(51 \AA)/Fe(34 \AA)]$_{15}$ multilayer. 
The multilayer was sputtered onto a Si substrate using Nb buffer (100 \AA) 
and cap (30 \AA) layers. 
SQUID magnetometry and XMCD measurements show that the multilayer couples
antiferromagnetically at the Gd/Fe interfaces and have coercive fields 
$< 50$ Oe at 300 K. 
X-ray measurements were performed at sector 4 of the Advanced Photon Source 
at Argonne National Laboratory. 
Undulator radiation was monochromatized with double Si(111) crystals and 
its polarization converted from linear to circular with a diamond (111) 
quarter-wave plate operated in Bragg transmission geometry.\cite{sector4} 
The sample was placed in a $B=2.1$ kG field parallel to its surface and 
in the scattering plane. 
Specular magnetic reflectivity was measured at room temperature with a photon energy
near the Gd L$_2$ resonance (7929 eV) across multilayer Bragg peaks by
switching the helicity of the incident radiation at each scattering vector
$q_z = (4\pi/\lambda)\sin\theta$, with $\theta$ being the grazing incidence angle.

\begin{figure}
\epsfxsize=14cm
\centerline{\epsffile{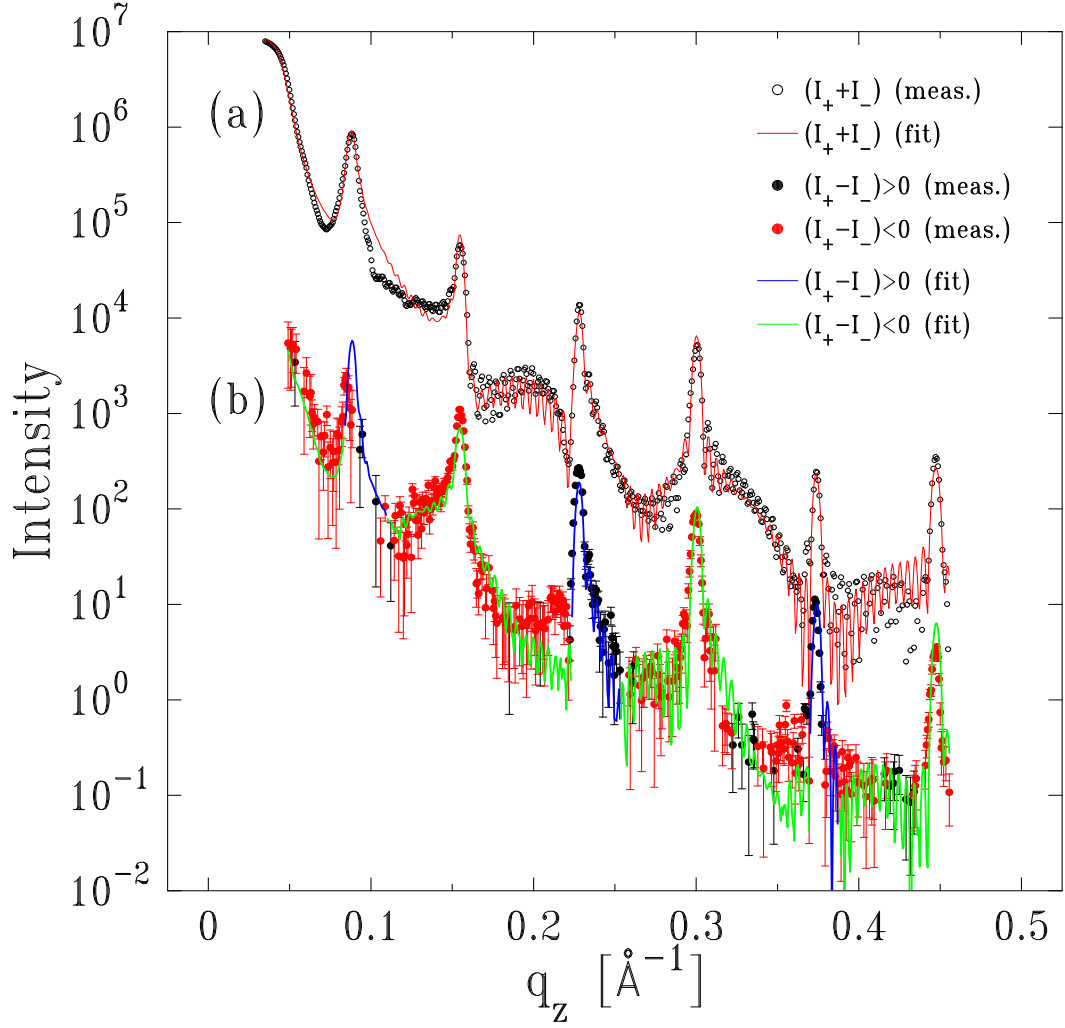}}
\caption{ \label{fig-fit}
$(I_++I_-)$ [(a)] and $(I_+-I_-)$ [(b)] intensities measured (symbols) from a
Fe(34 \AA)/[Gd(51 \AA)/Fe(34 \AA)]$_{15}$ multilayer near the Gd L$_2$-edge
(7929 eV). The lines represent the best theoretical fits with the model (B)
in Fig. \ref{mag-geo}. Note that the colors of symbols and lines in
$(I_+-I_-)$ intensities are different for opposite signs of the intensities.}
\end{figure}

Figure \ref{fig-fit} shows specular reflectivity curves obtained by adding
[(a), ($I_++I_-$)] and subtracting [(b), ($I_+-I_-$)] reflected intensities
for opposite helicites of the incoming x-rays. 
Symbols represent measurements and solid lines represent the fits calculated 
using Eq. (\ref{eq:sum_diff_inten}). 
From the fit for ($I_++I_-$) intensities, we obtained the layer thicknesses 
$d_{\rm Gd} = 50.74\pm0.09$ \AA~and $d_{\rm Fe} = 33.98\pm0.09$ \AA, 
and the roughness amplitudes of charge interfaces 
$\sigma_{c,{\rm Fe/Gd}} = 4.7\pm0.1$ \AA~and
$\sigma_{c,{\rm Gd/Fe}} = 3.6\pm0.1$ \AA. 
From the fit for ($I_+-I_-$) intensities, we found that the Gd layers were 
fully magnetized only near the Gd/Fe interfaces at room temperature, 
which is above the bulk $T_c$ of Gd.
This magnetization is induced by a strong antiferromagnetic exchange
interaction with the magnetically ordered Fe layers.\cite{camley} 
From the best fit, the thickness of the ferromagnetic Gd layer was estimated to be
$4.5\pm0.3$ \AA, which is consistent with our previous work.\cite{haskel01}
Magnetic roughness amplitudes for Gd/Fe (Fe/Gd) and
Gd-ferromagnetic/Gd-paramagnetic interfaces were estimated to be $4.2\pm0.1$
\AA~and $4.6\pm0.1$ \AA, respectively.


\section{conclusions}

The formulae for x-ray resonant magnetic specular reflectivity have been
derived for both single and multiple interfaces using the self-consistent
method in the framework of the distorted-wave Born approximation (DWBA). 
For this purpose, we have defined a structural and a magnetic interface
to represent the actual interfaces.
The well-known Nevot-Croce expression for the x-ray specular reflectivity from a
rough surface has been generalized and examined for the case of a
magnetically rough surface. 
The formalism has been generalized to the case of
multiple interfaces, as in the case of thin films or multilayers. 
Numerical illustrations have been given for typical examples of each 
of these systems and compared with the experimental data from a Gd/Fe multilayer. 
We have also presented the explicit expressions in the small-angle approximation,
which are readily applicable to transition-metal and rare-earth L-edge resonant
magnetic reflectivities.
The code for the calculations in this paper is also available in C language
by emailing to D.R.L. (\url{drlee@aps.anl.gov}).


\acknowledgments

Work at Argonne is supported by the U.S. DOE, Office of Basic Energy Sciences,
under Contract No. W-31-109-Eng-38. 

\appendix


\section{Explicit expressions for $R^{(0)}_{\nu\mu}$, $T^{(0)}_{j\mu}$ using
                        $2\times 2$ matrix formulae}

To calculate the explicit expressions for $R^{(0)}_{\nu\mu}$ and
$T^{(0)}_{j\mu}$ in Eq. (\ref{eq:E_ki}), we follow Stepanov and Sinha's
approach\cite{stepanov} developed for magnetic resonant reflections from 
ideally smooth interfaces. 
The electric field ${\bf E}_{z<0}({\bf r})$ inside
the magnetic medium with a dielectric susceptibility tensor given by 
Eq. (\ref{eq:chi_12}) can be represented as
\bb\label{eq:E_field}
{\bf E}_{z<0}({\bf r}) = {\bf E} e^{-i k_0 u z + ik_0 \cos\theta_i x},
\ee
where $\theta_i$ is the incidence angle, as shown in Fig. 1. 
The parameter $u$ can be a complex number due to absorption or total reflection. 
Substituting this in the wave equation Eq. (\ref{eq:wave_eq}), we obtain
\bb\label{eq:wave_eq1}
\sum_{\beta} \Bigl[ (\sin^2\theta_i - u^2) \delta_{\alpha\beta}
        + n_{\alpha} n_{\beta}
	+ \chi_{\alpha\beta} \Bigr] E_{\beta} = 0,
\ee
where $n_{\alpha} = k_{\alpha}/k_0$, i.e., $n_x=\cos\theta_i$, $n_y=0$, and
$n_z = -u$.

If we consider the case where the magnetization vector is aligned along the
sample surface in the scattering plane, i.e., ${\bf M}\parallel {\bf\hat{x}}$ in
Fig. 1, the tensor $\chi_{\alpha\beta}$ of a resonant magnetic medium can be
written from Eq. (\ref{eq:chi_tot}) as
\bb\label{eq:chi_tensor}
\Bigl(\chi_{\alpha\beta}\Bigr)_{{\bf M}\parallel {\bf\hat{x}}} =
 \Bigl( \chi_1\delta_{\alpha\beta} - i B^{\prime}\sum_{\gamma}
                                                 \epsilon_{\alpha\beta\gamma}
 M_{\gamma} + C^{\prime} M_{\alpha}M_{\beta} \Bigr)_{{\bf M}\parallel {\bf\hat{x}}}
= \left( \begin{array}{ccc} \chi_1+ C^{\prime} & 0 & 0 \\
			 0 & \chi_1 & -i B^{\prime} \\
			0 & i B^{\prime} & \chi_1  \end{array} \right),
\ee
where
\bb\label{eq:BC_prime}
\chi_1 = -\frac{4\pi}{k_0^2} \rho_0(\vec{r}) r_0
+ \frac{4\pi}{k_0^2} A n_m(\vec{r}),~~
B^{\prime} = \frac{4\pi}{k_0^2} B n_m(\vec{r}) M_x,~~
C^{\prime} = \frac{4\pi}{k_0^2} C n_m(\vec{r}) M_x^2.
\ee
Assuming that the incidence angle $\theta_i$ is small
($\sin\theta_i\approx\theta_i\ll 1$ and $n_x = \cos\theta_i\approx 1$) 
and even at the resonance
$\chi_{\alpha\beta}$ remain small ($|\chi_{\alpha\beta}|\ll 1$), and
inserting Eq. (\ref{eq:chi_tensor}) into Eq. (\ref{eq:wave_eq1}), 
the dispersion equation for a nontrivial
solution of Eq. (\ref{eq:wave_eq1}) can  be then approximated by
\bb\label{eq:dispersion_eq}
\left|\left| \begin{array}{ccc}
		1 & 0 & -u \\
		0 & \theta_i^2 + \chi_1 -u^2 & -i B^{\prime} \\
		-u & i B^{\prime} & \theta_i^2+ \chi_1
\end{array} \right|\right| = 0,
\ee
and the respective roots are $u^{(1,2,3,4)} = \pm\sqrt{\theta_i^2+\chi_1\pm
B^{\prime}}$. 
Two roots of these $u^{(j)}$'s with ${\rm Im}[u^{(1,2)}]>0$ and
the other two roots with ${\rm Im}[u^{(3,4)}]<0$ correspond to transmitted
and reflected waves in the medium, respectively. 
For each of the waves Eqs. (\ref{eq:wave_eq1}) and (\ref{eq:dispersion_eq}) give 
$(j=1,...,4)$
\bb\label{eq:wave_solution}
E_z^{(j)} = \frac{\theta_i^2 + \chi_1-u^{(j) 2}}{i B^{\prime}} E_y^{(j)},~~
E_x^{(j)} = u^{(j)}\frac{\theta_i^2 + \chi_1-u^{(j) 2}}{i B^{\prime}}
                                                       E_y^{(j)},~~
E_y^{(j)} = E_{\sigma}^{(j)}.
\ee
And if we denote
\bb\label{eq:u_pm}
u^{(1)} &=& \sqrt{ \theta_i^2+\chi_1+B^{\prime}} \equiv u_+,~~
u^{(2)} = \sqrt{ \theta_i^2+\chi_1-B^{\prime}} \equiv u_-,~~ \nonumber \\
u^{(3)} &=& -u_+,~~
u^{(4)} = - u_-,
\ee
we may then write
\bb\label{eq:wave_1234}
E_z^{(1)} &=& i E_{\sigma}^{(1)},~~
E_z^{(2)} = -i E_{\sigma}^{(2)},~~
E_z^{(3)} = i E_{\sigma}^{(3)},~~
E_z^{(4)} = -i E_{\sigma}^{(4)},~~\nonumber \\
E_x^{(j)} &=& u^{(j)}E_z^{(j)}~~~(j=1,...,4).
\ee
Since $|u^{(j)}|\ll 1$, $E_x^{(j)}$ can be neglected, then the
polarizations of the waves ${\bf\hat{e}}^{(j)}$ in the magnetic resonant medium
can be reduced to the circular polarizations
\bb\label{eq:pol_1234}
{\bf\hat{e}}^{(j)} &\approx& E_y^{(j)}{\bf\hat{e}}_{\sigma} + 
		E_z^{(j)} {\bf\hat{e}}_{\pi}~~~
               ({\bf\hat{y}} = {\bf\hat{e}}_{\sigma},
                  ~~{\bf\hat{z}}\approx {\bf\hat{e}}_{\pi}),~~~ \nonumber \\
{\bf\hat{e}}^{(1)} &=& {\bf\hat{e}}_{\sigma} + i {\bf\hat{e}}_{\pi}  
		= {\bf\hat{e}}^{(3)},~~~
{\bf\hat{e}}^{(2)} = {\bf\hat{e}}_{\sigma} - i {\bf\hat{e}}_{\pi}  
		= {\bf\hat{e}}^{(4)}.
\ee
If the wave field ${\bf E}_{z>0}({\bf r})$ with the incident and specularly
reflected waves inside the non-magnetic(isotropic) medium can be represented
as
\bb\label{eq:E_field_nonmag}
{\bf E}_{z>0}({\bf r}) = \left({\bf E}_0 e^{-i k_0 u_0 z} + {\bf E}_R
e^{ik_0 u_0 z} \right)
e^{i k_0 \cos\theta_i x},~~\left(u_0=\sqrt{ \theta_i^2+\chi_0 }\right),
\ee
the boundary conditions for the waves, ${\bf E}_{z>0}({\bf r})$ and
${\bf E}_{z<0}({\bf r})$ in Eqs. (\ref{eq:E_field}) and (\ref{eq:E_field_nonmag}) 
must be satisfied for the lateral components ${\bf E}_{\parallel}$ and 
${\bf H}_{\parallel}$ of
electric fields and magnetic fields, respectively. 
Since ${\bf H}\propto[{\bf\hat{k}}\times{\bf E}]$, this gives
\bb\label{eq:boundary_condi}
u_0 E_{0\pi}  - u_0 E_{R\pi} &=& \sum_j E_x^{(j)} \\ \nonumber
 E_{0\sigma}  +  E_{R\sigma} &=& \sum_j E_y^{(j)} \\ \nonumber
u_0 E_{0\sigma}  - u_0 E_{R\sigma} &=& u^{(j)} \sum_j E_y^{(j)} \\ \nonumber
 E_{0\pi}  + E_{R\pi} &=& \sum_j (u^{(j)} E_x^{(j)} + n_x  E_z^{(j)}) \approx
\sum_j E_z^{(j)},
\ee
where the approximation in the last equation was obtained by $|u^{(j)}|\ll 1$
and $n_x\approx 1$. 
Using Eqs. (\ref{eq:wave_solution})-(\ref{eq:wave_1234}), 
the above equations can be expressed in the $4\times 4$ matrix form
\bb\label{eq:bc_matrix}
\left( \begin{array}{cccc} 1 & 0 & 1 & 0 \\ 0 & 1 & 0 & 1\\
              u_0 & 0 & - u_0 & 0 \\
		0 & u_0 & 0 & - u_0 \end{array} \right)
\left( \begin{array}{c} E_{0\sigma} \\ E_{0\pi}\\E_{R\sigma}\\ E_{R\pi}
						  \end{array} \right)
= \left( \begin{array}{cccc} 1 & 1 & 1 & 1 \\ i & -i & i & -i\\
                                             u_+ & u_- & -u_+ & -u_- \\
		i u_+ & -i u_- & -i u_+ & i u_- \end{array} \right)
\left( \begin{array}{c} E_{\sigma}^{(1)} \\
    E_{\sigma}^{(2)}\\E_{\sigma}^{(3)}\\ E_{\sigma}^{(4)} \end{array} \right).
\ee
Representing the waves as the vectors $T_0 = (E_{0\sigma}, E_{0\pi})$, $R_0 =
(E_{R\sigma}, E_{R\pi})$, $T_1 = (E_{\sigma}^{(1)}, E_{\sigma}^{(2)})$, and
$R_1 = (E_{\sigma}^{(3)}, E_{\sigma}^{(4)})$, the $4\times4$ matrices in Eq.
(\ref{eq:bc_matrix}) can be reduced into four $2\times 2$ blocks
\bb\label{eq:X_mat}
\left( \begin{array}{c} T_0 \\ R_0 \end{array} \right) =
\left(
       \begin{array}{cc} X^{tt} & X^{tr} \\ X^{rt} & X^{rr} \end{array}
\right)
\left( \begin{array}{c} T_1 \\ R_1 \end{array} \right),
\ee
where $X^{tt}$, $X^{tr}$, $X^{rt}$, $X^{rr}$ can be obtained by multiplying
the inverse of the $4\times4$ matrix at the left side of 
Eq. (\ref{eq:bc_matrix}) onto the both sides. 
Since the reflected waves inside the medium vanish for a single
surface, ${\bf E}^{(3)}={\bf E}^{(4)}=0$ [i.e., $R_1 = (0,0)$], the
``unknown'' waves $R_0$ and $T_1$ in Eq. (\ref{eq:X_mat}) can be expressed via the
``known'' waves $T_0$ and $R_1$ as
\bb\label{eq:M_mat_def}
\left( \begin{array}{c} T_1 \\ R_0 \end{array} \right) =
\left(
       \begin{array}{cc} M^{tt} & M^{tr} \\ M^{rt} & M^{rr} \end{array}
\right)
\left( \begin{array}{c} T_0 \\ R_1 \end{array} \right),
\ee
where
\bb\label{eq:M_matrices}
M^{tt} &=& (X^{tt})^{-1},~~M^{tr}= - (X^{tt})^{-1}X^{tr},~~ \nonumber \\
M^{rt} &=& X^{rt}(X^{tt})^{-1},~~M^{rr}=X^{rr}-X^{rt}(X^{tt})^{-1}X^{tr}.
\ee
From Eqs. (\ref{eq:bc_matrix})-(\ref{eq:M_matrices}), 
the explicit expressions for $M^{pq}_{n\rightarrow r}$ matrices are given by
\bb\label{eq:M_mat_explicit}
M^{tt}_{n\rightarrow r} &=& \left( \begin{array}{cc} \frac{u_0}{u_0+u_+} & -i
                                                       \frac{u_0}{u_0+u_+} \\
           \frac{u_0}{u_0+u_-} & i \frac{u_0}{u_0+u_-} \end{array} \right)
		= T^{(0)}_{j\mu}({\bf k}_i),
                                                                 \nonumber \\
M^{tr}_{n\rightarrow r} &=&
      \left( \begin{array}{cc}
        \frac{u_+-u_0}{u_0+u_+} & 0 \\
		0               & \frac{u_--u_0}{u_0+u_-}
      \end{array} \right),
                                                                 \nonumber \\
M^{rt}_{n\rightarrow r} &=&
\left( \begin{array}{cc} \frac{u_0^2-u_+u_-}{(u_0+u_+)(u_0+u_-)}
			& i \frac{u_0(u_+ - u_-)}{(u_0+u_+)(u_0+u_-)} \\
		-i \frac{u_0(u_+ - u_-)}{(u_0+u_+)(u_0+u_-)}
		& \frac{u_0^2-u_+u_-}{(u_0+u_+)(u_0+u_-)} \end{array} \right)
		= R^{(0)}_{\nu\mu} ({\bf k}_i),
								\nonumber \\
M^{rr}_{n\rightarrow r} &=& \left( \begin{array}{cc}
       \frac{2 u_+}{u_0+u_+} & \frac{2 u_-}{u_0+u_-} \\
      i\frac{2 u_+}{u_0+u_+} & -i \frac{2 u_-}{u_0+u_-}
                            \end{array} \right),
\ee
where the $ij$-elements of $M^{pq}$ matrices are defined by Fig. 10, and 
the subscript $n\rightarrow r$ represents the incidence from a
nonmagnetic medium into a resonant magnetic one.  
From the definition of $M^{pq}$ matrices in Eq. (\ref{eq:M_mat_def}),
$R^{(0)}_{\nu\mu} ({\bf k}_i)$ and $T^{(0)}_{j\mu}({\bf k}_i)$ correspond to
$M^{rt}_{n\rightarrow r}$ and $M^{tt}_{n\rightarrow r}$, respectively.
For the time-reversed waves incident with vector $(-{\bf k}_f)$, 
scattering angle $\theta_f$, and
polarization $\nu$, $M^{pq}_{n\rightarrow r}(-{\bf k}_f)$ matrices are same
as the case of $({\bf k}_i, \mu)$ but replacing $i$ by $(-i)$ 
in Eq. (\ref{eq:M_mat_explicit}), i.e.,
\bb\label{eq:time_reversal_M}
M^{pq}_{n\rightarrow r}(-{\bf k}_f) &=& M^{pq}_{n\rightarrow r}
	({\bf k}_i ; i\leftrightarrow -i),~~(pq=tt,tr,rt,rr), \nonumber \\
T^{(0)}_{j\mu}(-{\bf k}_f) &=& T^{(0)}_{j\mu}({\bf k}_i; i\leftrightarrow -i),~~
R^{(0)}_{\nu\mu} (-{\bf k}_f) = R^{(0)}_{\nu\mu} ({\bf k}_i; i\leftrightarrow -i).
\ee

For completeness, let us now consider the reverse case where a wave is incident
``from'' a magnetic (anisotropic) medium with $\chi_{\alpha\beta} = \chi_1
\delta_{\alpha\beta} + \chi_{\alpha\beta}^{(2)}$ ``into'' a nonmagnetic one
with $\chi_{\alpha\beta} = \chi_0 \delta_{\alpha\beta}$. 
The explicit forms of $M^{pq}_{r \rightarrow n}$ matrices can be evaluated 
by starting with reversing both sides in Eq. (\ref{eq:bc_matrix}) 
and representing the waves as $T_0 = (E_{\sigma}^{(1)},
E_{\sigma}^{(2)})$, $R_0 = (E_{\sigma}^{(3)}, E_{\sigma}^{(4)})$, $T_1 =
(E_{0\sigma}, E_{0\pi})$, and $R_1 = (E_{R\sigma}, E_{R\pi})$ 
in Eq. (\ref{eq:X_mat}).
Then, $M^{pq}_{r\rightarrow n}$ matrices can be obtained straightforwardly by
\bb\label{eq:M_rn_mat}
M^{tt}_{r\rightarrow n} &=& T^{(0)}_{\mu j}({\bf k}_i) = 
	M^{rr}_{n\rightarrow r},~~
M^{tr}_{r\rightarrow n} = M^{rt}_{n\rightarrow r}, \nonumber \\
M^{rt}_{r\rightarrow n} &=& R^{(0)}_{j j^{\prime}} ({\bf k}_i) =
	M^{tr}_{n\rightarrow r},~~
M^{rr}_{r\rightarrow n} = M^{tt}_{n\rightarrow r}, 
\ee
where the subscript $r\rightarrow n$ denotes the incidence from a resonant
magnetic medium into a nonmagnetic one.
In the same way as in Eq. (\ref{eq:time_reversal_M}), 
the $M^{pq}_{r\rightarrow n}(-{\bf k}_f)$
matrices for the time-reversed waves can be also obtained by replacing $i$ by $(-i)$
in Eq. (\ref{eq:M_rn_mat}).  

Finally, let us also consider the magnetic-magnetic (resonant-resonant) 
interface between upper
($\chi_{\alpha\beta,{\rm up}} = \chi_{{\rm up}}\delta_{\alpha\beta}
+\chi_{\alpha\beta,{\rm up}}^{(2)}$) and lower ($\chi_{\alpha\beta,{\rm dw}}
= \chi_{{\rm dw}}\delta_{\alpha\beta} +\chi_{\alpha\beta,{\rm dw}}^{(2)}$)
resonant magnetic layers.
By employing the $4\times 4$ matrices involving resonant magnetic
medium to both sides of Eq. (\ref{eq:bc_matrix}), the explicit expressions
of $M^{pq}_{r\rightarrow r}$ can be given by
\bb\label{eq:M_mat_explicit_rr}
M^{tt}_{r\rightarrow r} &=& \left( \begin{array}{cc} 
			\frac{2 u_+^{\rm up}}
				{u_+^{\rm dw}+u_+^{\rm up}} & 0 \\
           0 & \frac{2 u_-^{\rm up}}{u_-^{\rm dw}+u_-^{\rm up}}
			 \end{array} \right),
                                                                 \nonumber \\
M^{tr}_{r\rightarrow r} &=&
      \left( \begin{array}{cc}
        \frac{u_+^{\rm dw}-u_+^{\rm up}}
		{u_+^{\rm dw}+u_+^{\rm up}} & 0 \\
                0   &  \frac{u_-^{\rm dw}-u_-^{\rm up}} 
			 {u_-^{\rm dw}+u_-^{\rm up}}
      \end{array} \right), ~~~ 
M^{rt}_{r\rightarrow r} = -M^{tr}_{r\rightarrow r} 
                                                                 \nonumber \\
M^{rr}_{r\rightarrow r} &=& \left( \begin{array}{cc}
	             \frac{2 u_+^{\rm dw}}
                                {u_+^{\rm dw}+u_+^{\rm up}} & 0 \\
           0 & \frac{2 u_-^{\rm dw}}{u_-^{\rm dw}+u_-^{\rm up}}                                         \end{array} \right),
\ee
where $u_{\pm}^{\rm up,dw} = \sqrt{ \theta_i^2 + \chi_{{\rm up,dw}}
\pm B^{\prime}_{\rm up,dw}}$ and $B^{\prime}_{\rm up,dw}$ was
defined in Eq. (\ref{eq:BC_prime}).
Note that these $M^{pq}_{r\rightarrow r}$ matrices for the magnetic-magnetic interfaces
are applicable to the nonmagnetic-nonmagnetic 
(nonresonant-nonresonant) interfaces simply by setting 
$B^{\prime}_{\rm up,dw}$ to be zero.


\section{Evaluation of the matrix elements involving $\bm{\chi}^{(0)}$}

To evaluate the matrix element in Eqs. (\ref{eq:chi0_mat}) and 
(\ref{eq:chi0_mat_trans}),
we assume that $\chi_0 = 0$ in Eq. (\ref{eq:chi0}), 
i.e., the first nonmagnetic medium is vacuum. 
Then the matrix element in Eq. (\ref{eq:chi0_mat}) can be
evaluated from Eqs. (\ref{eq:Ei}) and (\ref{eq:E_kf}) as
\bb\label{eq:chi0_mat_explicit_1}
k_0^2<-{\bf k}^T_f, \nu | \bm{\chi}^{(0)} | {\bf E}^i_{\mu}({\bf r})>
&=& {\cal A} k_0^2\delta_{k_{ix}k_{fx}} \delta_{k_{iy}k_{fy}} \sum_j
                           T^{(0)}_{j\nu}(-{\bf k}_f) \nonumber \\
&\times& \sum_{\alpha\beta} e^{\ast}_{j\alpha}
   (\chi_1 \delta_{\alpha\beta} + \chi_{\alpha\beta}^{(2)}) e_{\mu\beta}
      \int^{0}_{-\infty} dz e^{-i ( k^t_{fz}(j)-k_{iz})z },      \nonumber \\
&=& i {\cal A} k_0^2\delta_{k_{ix}k_{fx}} \delta_{k_{iy}k_{fy}}         \nonumber \\
&\times& \sum_j \frac{T^{(0)}_{j\nu}(-{\bf k}_f)}
                        { k^t_{fz}(j)-k_{iz} }
     \sum_{\alpha\beta} e^{\ast}_{j\alpha}
      (\chi_1 \delta_{\alpha\beta} + \chi_{\alpha\beta}^{(2)}) e_{\mu\beta}.
\ee

In order to evaluate the explicit expression for the above equation, 
let us now consider the case where the incidence angle $\theta_i$ is small and
${\bf M}\parallel{\bf\hat{x}}$, as discussed in Appendix A. 
In this case, ${\bf\hat{e}}_j = {\bf\hat{e}}_{\sigma}\pm i 
{\bf\hat{e}}_{\pi}$ and $k^t_{fz}(j) =
k_0 u_{\pm}$, where the upper and lower signs correspond to $j=1$ and 2,
respectively, and $k_{iz} = -k_0 u_0$. 
From Eqs. (\ref{eq:chi_12}), (\ref{eq:BC_prime}), and (\ref{eq:u_pm}), 
the polarization-dependent terms are evaluated by
\bb\label{eq:pol_term}
\sum_{\alpha\beta} e^{\ast}_{j\alpha} 
	(\chi_1 \delta_{\alpha\beta} + \chi_{\alpha\beta}^{(2)}) e_{\mu\beta} 
	&=& \chi_1 ({\bf\hat{e}}^{\ast}_j \cdot {\bf\hat{e}}_{\mu}) 
	+ ({\bf\hat{e}}^{\ast}_j \times {\bf\hat{e}}_{\mu})_x 
		\chi^{(2)}_{{\bf M}\parallel{\bf\hat{x}}} \nonumber \\
&=& \left\{ \begin{array}{cc} 
	\chi_1 + i \chi^{(2)} = u_+^2-u_0^2, 
	& {\rm for~} j=1, \mu=\sigma \\
	\chi_1 - i \chi^{(2)} = u_-^2-u_0^2, 
	& {\rm for~} j=2, \mu=\sigma \\
	-i\chi_1 + \chi^{(2)} = -i(u_+^2-u_0^2), 
	& {\rm for~} j=1, \mu=\pi \\
	i\chi_1 + \chi^{(2)} = i(u_-^2-u_0^2), 
	& {\rm for~} j=2, \mu=\pi 
                \end{array} \right.,  
\ee
where $u_0 = \theta_i$ when $\chi_0 = 0$ in Eq. (\ref{eq:E_field_nonmag}).
The explicit form of $2\times 2$ matrix $T^{(0)}_{j\nu}(-{\bf k}_f)$ 
can be obtained from $M^{tt}_{n\rightarrow r}$ in Eq. (\ref{eq:M_mat_explicit}) 
by replacing $i$ by $(-i)$. 
Then, the matrix element in Eq. (\ref{eq:chi0_mat}) can be expressed by 
$2\times 2$ matrix in terms of the polarizations of incident and outgoing beams, 
$\mu$ and $\nu$, as follows:
\bb\label{eq:chi0_mat_explicit_2}
k_0^2<-{\bf k}^T_f, \nu | \bm{\chi}^{(0)} | {\bf E}^i_{\mu}({\bf r})>
&=& i{\cal A} k_0^2\delta_{k_{ix}k_{fx}} \delta_{k_{iy}k_{fy}} \frac{2 u_0}{k_0
                                         (u_++u_0)(u_-+u_0)} \nonumber \\
&\times&
\left( \begin{array}{cc} u_+u_- - u_0^2 & -i u_0(u_+ - u_-) \\
                        i u_0(u_+ - u_-) & u_+u_- - u_0^2
                \end{array} \right),  \nonumber \\
&=& 2 i {\cal A} k_{iz} R^{(0)}_{\nu\mu}({\bf k}_i) \delta_{k_{ix}k_{fx}}
                                              \delta_{k_{iy}k_{fy}},
\ee
where $R^{(0)}_{\nu\mu}({\bf k}_i)$ corresponds to $M^{rt}_{n\rightarrow r}$ in
Eq. (\ref{eq:M_mat_explicit}).
Without loss of generality the final result in Eq. (\ref{eq:chi0_mat_explicit_2})
is applicable for the case with $\chi_0\neq 0$ although the calculation for 
$z>0$ should be included in 
Eqs. (\ref{eq:chi0_mat_explicit_1})-(\ref{eq:chi0_mat_explicit_2}). 

For the transmission coefficient, the matrix element in Eq. (\ref{eq:chi0_mat_trans}) 
for $\chi_0 = 0$ can be also 
evaluated from Eqs. (\ref{eq:E_ki_trans}) and (\ref{eq:E_kf_trans}) as
\bb\label{eq:chi0_mat_trans_explicit_1}
k_0^2<-{\bf k}^T_f, j | \bm{\chi}^{(0)} | {\bf k}_i, \mu>
&=& {\cal A} k_0^2\delta_{k_{ix}k_{fx}} \delta_{k_{iy}k_{fy}} \sum_{\nu}
                                   T^{(0)}_{\nu j}(-{\bf k}_f)
     \sum_{j^{\prime}} T^{(0)}_{j^{\prime}\mu}({\bf k}_i)        \nonumber \\
&\times& \sum_{\alpha\beta} e^{\ast}_{\nu\alpha}
 (\chi_1 \delta_{\alpha\beta} + \chi_{\alpha\beta}^{(2)})
                 e_{j^{\prime}\beta}
                   \int^{0}_{-\infty} dz
                    e^{-i (- k_{fz}-k^t_{iz}(j^{\prime}))z },
                                                                 \nonumber \\
&=& i {\cal A} k_0^2\delta_{k_{ix}k_{fx}} \delta_{k_{iy}k_{fy}}         \nonumber \\
&\times& \sum_{\nu,j^{\prime}} \frac{T^{(0)}_{\nu j}(-{\bf k}_f)
                                 T^{(0)}_{j^{\prime}\mu}({\bf k}_i)}
					{- k_{fz}-k^t_{iz}(j^{\prime})}
	 \sum_{\alpha\beta} e^{\ast}_{\nu\alpha}
                (\chi_1 \delta_{\alpha\beta} + \chi_{\alpha\beta}^{(2)})
                             e_{j^{\prime}\beta},
\ee
where the vector field ${\bf E}({\bf k}_i, \mu)$ in Eq. (\ref{eq:E_ki_trans}) 
has been used for the state $|{\bf k}_i,\mu>$ instead of the ``pure'' incoming wave
${\bf E}^i_{\mu}({\bf r})$ in Eq. (\ref{eq:Ei}). 
Similarly to the reflection coefficient 
in Eqs. (\ref{eq:chi0_mat_explicit_1})-(\ref{eq:chi0_mat_explicit_2}), 
the matrix element in Eq. (\ref{eq:chi0_mat_trans}) can be expressed by 
a $2\times 2$ matrix in terms of
the polarizations of incident and transmitted beams, $\mu$ and $j$, as follows:
\bb\label{eq:chi0_mat_trans_explicit_2}
k_0^2<-{\bf k}^T_f, j | \bm{\chi}^{(0)} | {\bf k}_i, \mu>
&=& i {\cal A} k_0^2\delta_{k_{ix}k_{fx}} \delta_{k_{iy}k_{fy}}  \frac{4}{k_0}
\left( \begin{array}{cc}
         u_+ \frac{u_0}{u_+ + u_0} & u_+ \frac{ -i u_0}{u_+ + u_0} \\
         u_- \frac{u_0}{u_- + u_0} & u_- \frac{ i u_0}{u_- + u_0}
       \end{array} \right),                                      \nonumber \\
&=& 4 i {\cal A} k^t_{iz}(j) \delta_{k_{ix}k_{fx}} \delta_{k_{iy}k_{fy}}
\left( \begin{array}{cc}
         \frac{u_0}{u_+ + u_0} & \frac{ -i u_0}{u_+ + u_0} \\
         \frac{u_0}{u_- + u_0} & \frac{ i u_0}{u_- + u_0}
        \end{array} \right),                                      \nonumber \\
&=& 4 i {\cal A} k^t_{iz}(j) T^{(0)}_{j\mu}({\bf k}_i)
                          \delta_{k_{ix}k_{fx}} \delta_{k_{iy}k_{fy}},
\ee
where $T^{(0)}_{j\mu}({\bf k}_i)$ corresponds to $M^{tt}_{n\rightarrow r}$ in
Eq. (\ref{eq:M_mat_explicit}).
Again, the final result in Eq. (\ref{eq:chi0_mat_trans_explicit_2}) 
is applicable for the case with $\chi_0\neq 0$ without loss of generality.


\section{Explicit expressions for rough-interface $\tilde{M}^{pq}$ matrices}

For the interface between upper nonmagnetic ($\chi_{\alpha\beta}=\chi_0
\delta_{\alpha\beta}$) and lower resonant magnetic
($\chi_{\alpha\beta}=\chi_1 \delta_{\alpha\beta} + \chi^{(2)}_{\alpha\beta}$)
layers, the explicit expressions of the rough-interface
$\tilde{M}^{pq}_{n\rightarrow r}$ matrices can be given by
\bb\label{eq:M_RT_nr_rough}
\tilde{M}^{rt}_{n\rightarrow r} &=& {\bf R}_{n\rightarrow r}
= ({\bf I} - {\bf V}_{n\rightarrow r})^{-1}( {\bf R}^{(0)}_{n\rightarrow r}
	+ {\bf U}_{n\rightarrow r}),                             \nonumber \\
\tilde{M}^{tt}_{n\rightarrow r} &=& {\bf T}_{n\rightarrow r}
= ({\bf I} - {\bf V^{\prime}}_{n\rightarrow r})^{-1}
                                       {\bf T}^{(0)}_{n\rightarrow r},
\ee
where, from Eqs. (\ref{eq:ref_final}) and (\ref{eq:trans_final}),
\bb\label{eq:RUVT_nr_explicit}
{\bf R}^{(0)}_{n\rightarrow r} + {\bf U}_{n\rightarrow r}
&=& -\frac{1}{2}
  \left( \begin{array}{cc}
  \frac{D^{(+)}_1 + D^{(+)}_2}{(u_+ + u_0)^2} + 
                \frac{D^{(+)}_3 - D^{(+)}_4}{(u_- + u_0)^2} &
   -i\left( \frac{D^{(+)}_1 + D^{(+)}_2}{(u_+ + u_0)^2} 
		- \frac{D^{(+)}_3 - D^{(+)}_4}{(u_- + u_0)^2}
                                                                    \right) \\
     i\left( \frac{D^{(+)}_1 + D^{(+)}_2}{(u_+ + u_0)^2} 
		- \frac{D^{(+)}_3 - D^{(+)}_4}{(u_- + u_0)^2}
                                                                     \right) &
     \frac{D^{(+)}_1 + D^{(+)}_2}{(u_+ + u_0)^2} 
		+ \frac{D^{(+)}_3 - D^{(+)}_4}{(u_- + u_0)^2}
\end{array} \right),                                             \nonumber \\
{\bf I} - {\bf V}_{n\rightarrow r}
&=& \frac{1}{2}
  \left( \begin{array}{cc} \frac{D^{(-)}_1 + D^{(-)}_2}{u^2_+ - u_0^2}
	+ \frac{D^{(-)}_3 - D^{(-)}_4}{u_-^2 - u_0^2} &
        -i\left( \frac{D^{(-)}_1 + D^{(-)}_2}{u_+^2 - u_0^2}
		- \frac{D^{(-)}_3 - D^{(-)}_4}{u_-^2 - u_0^2} \right) \\
        i\left( \frac{D^{(-)}_1 + D^{(-)}_2}{u_+^2 - u_0^2}
		- \frac{D^{(-)}_3 - D^{(-)}_4}{u_-^2 - u_0^2} \right) &
                       \frac{D^{(-)}_1 + D^{(-)}_2}{u_+^2 + u_0^2}
	+ \frac{D^{(-)}_3 - D^{(-)}_4}{u_-^2 - u_0^2}
\end{array} \right),                                              \nonumber \\
{\bf I} - {\bf V^{\prime}}_{n\rightarrow r}
&=&  \left( \begin{array}{cc} \frac{D^{(-)}_1 +
                                           D^{(-)}_2}{u^2_+ - u_0^2} & 0 \\
			0 & \frac{D^{(-)}_3 - D^{(-)}_4}{u_-^2 - u_0^2}
		\end{array} \right), \nonumber \\
{\bf T}^{(0)}_{n\rightarrow r} &=&  M^{tt}_{n\rightarrow r}
= \left( \begin{array}{cc} \frac{u_0}{u_+ + u_0} & -i \frac{u_0}{u_+ + u_0} \\
			\frac{u_0}{u_- + u_0} & i \frac{u_0}{u_- + u_0}
		\end{array} \right),
\ee
and
\bb\label{eq:D_def}
D_1^{(\pm)} &=& (\chi_1 - \chi_0) 
		e^{-\frac{k_0^2}{2}(u_+\pm u_0)^2\sigma_c^2},~~
D_2^{(\pm)} = B^{\prime} 
		e^{-\frac{k_0^2}{2}(u_+\pm u_0)^2\sigma_m^2},~~     \nonumber \\
D_3^{(\pm)} &=& (\chi_1 - \chi_0) 
		e^{-\frac{k_0^2}{2}(u_-\pm u_0)^2\sigma_c^2},~~
D_4^{(\pm)} = B^{\prime} 
		e^{-\frac{k_0^2}{2}(u_-\pm u_0)^2\sigma_m^2}.       
\ee
Here, $(\chi_1 - \chi_0) = (u^2_+ + u^2_-)/2 - u_0^2$ and $B^{\prime} = (u^2_+
- u^2_-)/2$ can be used from $u_0 = \sqrt{\theta_i^2 + \chi_0}$ and
$u_{\pm}=\sqrt{\theta_i^2+\chi_1\pm B^{\prime}}$.

For the reversed interface between upper magnetic(resonant) and lower
nonmagnetic layers, $\tilde{M}^{pq}_{r \rightarrow n}$ matrices can be also
given by
\bb\label{eq:M_RT_rn_rough}
\tilde{M}^{rt}_{r\rightarrow n} &=& {\bf R}_{r\rightarrow n}
= ({\bf I} - {\bf V}_{r\rightarrow n})^{-1}( {\bf R}^{(0)}_{r\rightarrow n}
	+ {\bf U}_{r\rightarrow n}), \nonumber \\
\tilde{M}^{tt}_{r\rightarrow n} &=& {\bf T}_{r\rightarrow n}
= ({\bf I} - {\bf V^{\prime}}_{r\rightarrow n})^{-1}
                                             {\bf T}^{(0)}_{r\rightarrow n},
\ee
where
\bb\label{eq:RUVT_rn_explicit}
{\bf R}^{(0)}_{r\rightarrow n} + {\bf U}_{r\rightarrow n}
&=&  \left( \begin{array}{cc} \frac{D^{(+)}_1 + D^{(+)}_2}{(u_+ + u_0)^2} & 0 \\
			0 & \frac{D^{(+)}_3 - D^{(+)}_4}{(u_- + u_0)^2}
	\end{array} \right), \nonumber \\
{\bf I} - {\bf V}_{r\rightarrow n} &=&
                              {\bf I} - {\bf V^{\prime}}_{n\rightarrow r},~~~
{\bf I} - {\bf V^{\prime}}_{r\rightarrow n} = {\bf I} -
                                           {\bf V}_{n\rightarrow r},
                                                                 \nonumber \\
{\bf T}^{(0)}_{r\rightarrow n} &=&  M^{tt}_{r\rightarrow n}
= \left( \begin{array}{cc} \frac{2u_+}{u_+ + u_0} & \frac{2u_-}{u_- + u_0} \\
			i \frac{2 u_+}{u_+ + u_0} & -i \frac{2u_-}{u_- + u_0}
		\end{array} \right).
\ee
In the same way as Eq. (\ref{eq:M_rn_mat}), 
other $\tilde{M}^{pq}$ matrices can be given by
\bb\label{eq:otherM}
\tilde{M}^{tr}_{n\rightarrow r} = \tilde{M}^{rt}_{r\rightarrow n},~~
\tilde{M}^{rr}_{n\rightarrow r} = \tilde{M}^{tt}_{r\rightarrow n},~~
\tilde{M}^{tr}_{r\rightarrow n} = \tilde{M}^{rt}_{n\rightarrow r},~~
\tilde{M}^{rr}_{r\rightarrow n} = \tilde{M}^{tt}_{n\rightarrow r}.
\ee

Finally, for the magnetic-magnetic (resonant-resonant) interface between
upper resonant magnetic ($\chi_{\alpha\beta}^{\rm up} = 
\chi_1^{\rm up}\delta_{\alpha\beta}+\chi_{\alpha\beta}^{(2),{\rm up}}$)
and lower resonant magnetic ($\chi_{\alpha\beta}^{\rm dw}
= \chi_1^{\rm dw}\delta_{\alpha\beta} +\chi_{\alpha\beta}^{(2),{\rm dw}}$)
layers, $\tilde{M}^{pq}_{r \rightarrow n}$ matrices can be also
given by
\bb\label{eq:M_RT_rr_rough}
\tilde{M}^{rt}_{r\rightarrow r} &=& {\bf R}_{r\rightarrow r}
= ({\bf I} - {\bf V}_{r\rightarrow r})^{-1}( {\bf R}^{(0)}_{r\rightarrow r}
        + {\bf U}_{r\rightarrow r}), \nonumber \\
\tilde{M}^{tt}_{r\rightarrow r} &=& {\bf T}_{r\rightarrow r}
= ({\bf I} - {\bf V^{\prime}}_{r\rightarrow r})^{-1}
                                             {\bf T}^{(0)}_{r\rightarrow r},
\ee
where                   
\bb\label{eq:RUVT_rr_explicit}
{\bf R}^{(0)}_{r\rightarrow r} + {\bf U}_{r\rightarrow r}
&=&  - \left( \begin{array}{cc} 
		\frac{D^{(+)}_5 + D^{(+)}_6}
			{(u_+^{\rm dw} + u_+^{\rm up})^2} & 0 \\
                 0 & \frac{D^{(+)}_7 - D^{(+)}_8}
			{(u_-^{\rm dw} + u_-^{\rm up})^2}
        \end{array} \right), \nonumber \\
{\bf I} - {\bf V}_{r\rightarrow r} 
&=&  \left( \begin{array}{cc}
                \frac{D_5^{(-)} + D_6^{(-)}}
                        {(u_+^{\rm dw})^2 - (u_+^{\rm up})^2} & 0 \\
                 0 & \frac{D_7^{(-)} - D_8^{(-)}}
                        {(u_-^{\rm dw})^2 - (u_-^{\rm up})^2}
        \end{array} \right), 
\ee
and
\bb\label{eq:D_def_rr}
D_5^{(\pm)} &=& (\chi_1^{\rm dw} - \chi_1^{\rm up}) 
	e^{-\frac{k_0^2}{2}(u_+^{\rm dw}\pm u_+^{\rm up})^2\sigma_c^2},~~
D_6^{(\pm)} = (B^{\prime}_{\rm dw}-B^{\prime}_{\rm up})
	e^{-\frac{k_0^2}{2}(u_+^{\rm dw}\pm u_+^{\rm up})^2\sigma_m^2},~~
	                                           \nonumber \\
D_7^{(\pm)} &=& (\chi_1^{\rm dw} - \chi_1^{\rm up}) 
	e^{-\frac{k_0^2}{2}(u_-^{\rm dw}\pm u_-^{\rm up})^2\sigma_c^2},~~
D_8^{(\pm)} = (B^{\prime}_{\rm dw}-B^{\prime}_{\rm up})
	e^{-\frac{k_0^2}{2}(u_-^{\rm dw}\pm u_-^{\rm up})^2\sigma_m^2},~~
\ee  
and $({\bf I} - {\bf V^{\prime}}_{r\rightarrow r})$ corresponds to
$({\bf I} - {\bf V}_{r\rightarrow r})$ when switching the upper and
lower layers, and ${\bf T}^{(0)}_{r\rightarrow r}$ corresponds to
$M^{tt}_{r\rightarrow r}$ in Eq. (\ref{eq:M_mat_explicit_rr}).
Here, $(\chi_1^{\rm dw} - \chi_1^{\rm up}) = 
[(u_+^{\rm dw})^2 + (u_-^{\rm dw})^2]/2 
- [(u_+^{\rm up})^2 + (u_-^{\rm up})^2]/2 $ and 
$(B_{\rm dw}^{\prime}-B_{\rm up}^{\prime}) = 
[(u_+^{\rm dw})^2 - (u_-^{\rm dw})^2]/2
- [(u_+^{\rm up})^2 - (u_-^{\rm up})^2]/2 $
can be used from
$u_{\pm}^{\rm up,dw}=\sqrt{\theta_i^2+\chi_1^{\rm up,dw}\pm 
B^{\prime}_{\rm up,dw}}$.
In the same way as Eq. (\ref{eq:otherM}), two other matrices 
$\tilde{M}^{tr}_{r\rightarrow r}$ and $\tilde{M}^{rr}_{r\rightarrow r}$ 
can be also obtained from 
$\tilde{M}^{rt}_{r\rightarrow r}$ and $\tilde{M}^{tt}_{r\rightarrow r}$
in Eq. (\ref{eq:M_RT_rr_rough}), respectively, by switching the upper
and lower layers.
We should mention again that these rough-interface $\tilde{M}^{pq}$
matrices for the magnetic-magnetic (resonant-resonant) interfaces
can be reduced to the cases for the nonmagnetic-nonmagnetic 
(nonresonant-nonresonant) interfaces by setting 
$B^{\prime}_{\rm up,dw}$ to be zero.


\section{Solutions of self-consistent matrix equations for nonmagnetic interfaces}

For nonmagnetic interfaces $(|{\bf M}|=0)$ and $\sigma\rightarrow\sigma$
polarization, simply
\bb\label{eq:u_pm_nonmag}
u_+=u_-=\sqrt{\theta_i^2+\chi_1} = |k_z^t|/k_0.
\ee
Inserting this in Eq. (\ref{eq:M_mat_explicit}) modified for $(-{\bf k}_f)$ 
and using $\chi_1-\chi_0=(|k_z^t|^2-|k_z|^2)/k_0^2$, the self-consistent solution 
for the reflection coefficient $({\bf k}_f={\bf k}_i^r$ and $\theta_i=\theta_f)$ in
Eq. (\ref{eq:ref_final}) can be reduced to a scalar as
\bb\label{eq:ref_final_scalar}
R=(1-V)^{-1}(R^{(0)} + U),
\ee
where
\bb\label{eq:RUV_scalar}
U &=& \frac{|k_z|-|k_z^t|}{|k_z|+|k_z^t|}\left(
             e^{-\frac{1}{2}(|k_z|+|k_z^t|)^2\sigma_c^2} -1\right),
                                                                \\ \nonumber
R^{(0)} &=& \frac{|k_z|-|k_z^t|}{|k_z|+|k_z^t|},                \\ \nonumber
V &=& 1- e^{-\frac{1}{2}(|k_z|-|k_z^t|)^2\sigma_c^2},
\ee
and $k_{fz}^t = |k_z^t|$, and $k_{iz}=-k_{iz}^r=-|k_z|$. 
Then, we obtain
\bb\label{eq:NC_scalar}
R = \frac{|k_z|-|k_z^t|}{|k_z|+|k_z^t|}
     e^{-\frac{1}{2}(|k_z|+|k_z^t|)^2\sigma_c^2}
     e^{+\frac{1}{2}(|k_z|-|k_z^t|)^2\sigma_c^2}
= R^{(0)} e^{-2|k_z||k_z^t|\sigma_c^2},
\ee
which is consistent with the Nevot-Croce form.\cite{nevot}

Similarly, the self-consistent solution for the transmission coefficient in
Eq. (\ref{eq:trans_final}) can be reduced into a scalar as
\bb\label{eq:trans_final_scalar}
T_{(\sigma\sigma)} = \sum_{j=1,2} T_{j\sigma}
                   = \sum_{j=1,2}(1-V^{\prime})^{-1} T^{(0)}_{j\sigma},
\ee
where
\bb\label{eq:TV_scalar}
V^{\prime} &=& 1 - e^{-\frac{1}{2}\left( |k_z|-|k_z^t|\right)^2 \sigma_c^2},
								 \nonumber \\
T^{(0)}_{1\sigma} &=& T^{(0)}_{2\sigma} = \frac{|k_z|}{|k_z|+|k_z^t|}.
\ee
Then,
\bb\label{eq:VV_scalar}
T = \frac{2|k_z|}{|k_z|+|k_z^t|} e^{\frac{1}{2}\left( |k_z|-|k_z^t|\right)^2
\sigma_c^2} = T^{(0)} e^{\frac{1}{2}\left( |k_z|-|k_z^t|\right)^2 \sigma_c^2},
\ee
which is consistent with the Vidal-Vincent form.\cite{vidal}


\section{Recursive $2\times 2$ matrix formulae for multiple interfaces}

For multiple interfaces, additional phase differences between different
interfaces should be taken into account to extend the results for a single
interface in Appendix A. 
Following Ref.~\onlinecite{stepanov}, $M^{pq}_{n+1}$
matrices for the $n$-th interface between $n$- and $(n+1)$-th layers can be
modified from Eq. (\ref{eq:M_matrices}) as
\bb\label{eq:MF_mat}
M^{tt}_{n+1} = M^{tt} F_n^{-1},~~
M^{tr}_{n+1} = M^{tr},~~
M^{rt}_{n+1} = F_n^{-1} M^{rt} F_n^{-1},~~
M^{rr}_{n+1} = F_n^{-1} M^{rr},
\ee
where $M^{pq}$ are the $2\times 2$ matrices obtained for a single smooth interface 
in Appendix A, depending on whether the upper and lower layers on the $n$-th interface
are nonmagnetic or magnetic ones, respectively,
\bb\label{eq:F_mat}
F_n =
\left( \begin{array}{cc} e^{-ik_0 u_{+,n} d_n} & 0 \\
                                             0 & e^{-ik_0 u_{-,n} d_n}
\end{array} \right),
\ee
and $u_{\pm,n}$ and $d_n$ represent the refracted angle defined in 
Eq. (\ref{eq:u_pm}) and the thickness of the $n$-th (upper) layer, respectively. 
For nonmagnetic layers, $u_{\pm,n}$ reduces to $u_{0,n}$ 
in Eq. (\ref{eq:E_field_nonmag}). 
$R_n$ and $T_n$ are the vectors $(R_{n,1}, R_{n,2})$ and 
$(T_{n,1}, T_{n,2})$ representing the two waves reflected and transmitted, 
respectively, at the top of the $n$-th layer.
(In Ref.~\onlinecite{stepanov}, they are defined at the bottom of the $n$-th layer.)

Introducing $W^{pq}_{n}$ matrices following Ref.~\onlinecite{stepanov}, which
are defined by
\bb\label{eq:W_mat_def}
\left(  \begin{array}{c} T_n \\ R_0 \end{array} \right)
= \left( \begin{array}{cc} W^{tt}_n & W^{tr}_n \\ W^{rt}_n & W^{rr}_n
         \end{array} \right)
         \left(  \begin{array}{c} T_0 \\ R_n
         \end{array} \right),
\ee
and using the recursion formulae involving $M^{pq}_{n+1}$ matrices at the
$n$-th interface, i.e.,
\bb\label{eq:M_mat_def_ml}
\left(  \begin{array}{c} T_{n+1} \\ R_n \end{array} \right)
= \left( \begin{array}{cc} M^{tt}_{n+1} & M^{tr}_{n+1} \\
                           M^{rt}_{n+1} & M^{rr}_{n+1}
         \end{array} \right)
\left(  \begin{array}{c} T_n \\ R_{n+1} \end{array} \right),
\ee
yields the following recursion formulae for $W^{pq}_n$ matrices:
\bb\label{eq:WM_mat_ml}
W^{tt}_{n+1} &=& A_n W^{tt}_n, \nonumber \\
W^{tr}_{n+1} &=& M^{tr}_{n+1} + A_n W^{tr}_n M^{rr}_{n+1}, \nonumber \\
W^{rt}_{n+1} &=& W^{rt}_n + B_n M^{rt}_{n+1} W^{tt}_n, \nonumber \\
W^{rr}_{n+1} &=& B_n M^{rr}_{n+1},
\ee
where $A_n$ and $B_n$ are defined by
\bb\label{eq:AB_mat_ml}
A_n = M^{tt}_{n+1} \left( 1 - W^{rt}_n M^{rt}_{n+1}\right)^{-1}, \nonumber \\
B_n = W^{rr}_n \left( 1 - M^{rt}_{n+1} W^{tr}_n\right)^{-1}.
\ee
Here $W^{rt}_N$ determines the reflectivity of the whole multilayer, $R_0 =
W^{rt}_N T_0$ ($R_N = 0$), from Eq. (\ref{eq:W_mat_def}).

Finally, the field amplitudes $T_n$, $R_n$ inside the layers can be obtained
from Eqs. (\ref{eq:W_mat_def})-(\ref{eq:AB_mat_ml}) by
\bb\label{eq:RT_recursive}
R_n &=& \left( 1 - M^{rt}_{n+1}W^{tr}_n \right)^{-1}
        \left( M^{rr}_{n+1} R_{n+1} + M^{rt}_{n+1} W^{tt}_n T_0 \right),
                                                                 \nonumber \\
T_n &=& W^{tt}_n T_0 + W^{tr}_n R_n,
\ee
which must be progressively applied to all the layers starting at the
multilayer substrate where $R_N = 0$.



\newpage

\newpage

\newpage

\newpage

\newpage

\newpage

\newpage

\newpage

\newpage

\newpage
\begin{figure}
\epsfxsize=8cm
\centerline{\epsffile{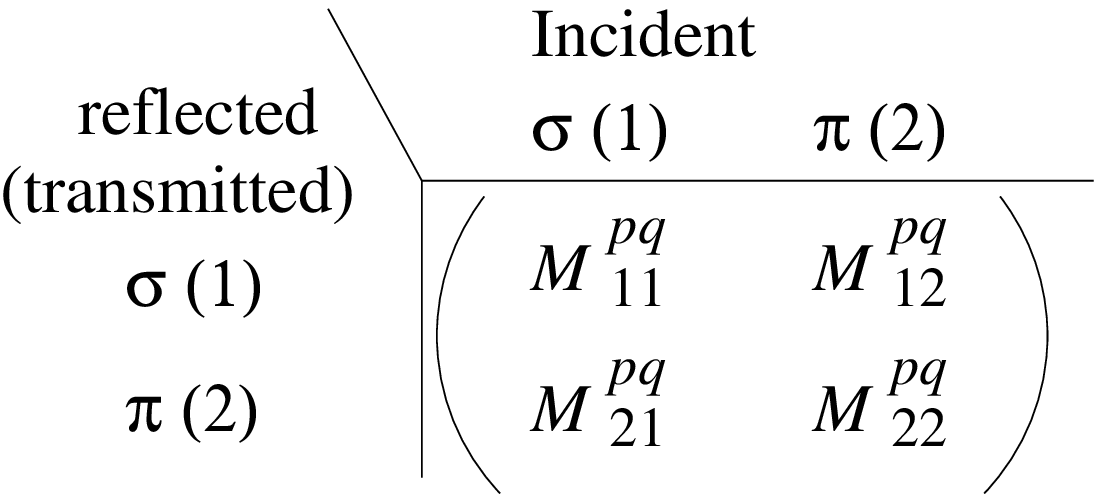}}
\caption{ \label{fig:M_matrix_def}
The representation chosen for the elements of $M^{pq}$ matrices with the 
polarization bases of the incident and reflected (or transmitted) waves.
The polarization basis is given by (${\bf\hat{e}}_{\sigma}$, ${\bf\hat{e}}_{\pi}$),
as shown in Fig. 1, for the waves in the nonmagnetic medium and
(${\bf\hat{e}}^{(1)}$, ${\bf\hat{e}}^{(2)}$), as defined in Appendix A,
for those in the resonant magnetic medium, respectively.      
}
\end{figure}

\end{document}